\documentclass[10pt, a4paper, twocolumn]{article}
\usepackage[left=2.75cm,right=2.75cm,top=4cm,bottom=4cm]{geometry}
\usepackage{fancyhdr}
\usepackage{pdfpages}
\usepackage[utf8]{inputenc}
\usepackage[english]{babel}
\usepackage{graphicx}
\usepackage{mathptmx}
\usepackage{amsmath}			%
\usepackage{amssymb}			
\usepackage{amsfonts}
\usepackage{graphicx}			
\usepackage{xcolor}				
\usepackage{siunitx}			
\usepackage{textcomp}           
\usepackage[sort&compress, numbers, square]{natbib}
\newcommand{\ie}{i.\,e.,\ }		
\newcommand{\eg}{e.\,g.,\ }		
\newcommand{\eqspace}{\:}	
\newcommand{\sref}[1]{section~\ref{#1}}

\newcommand{\fref}[1]{figure~\ref{#1}}
\newcommand{\Fref}[1]{Figure~\ref{#1}}

\newcommand{\rbrak}[1]{\left(#1\right)}						
\newcommand{\sbrak}[1]{\left[#1\right]}						
\newcommand{\cbrak}[1]{\left\{#1\right\}}					
\newcommand{\func}[2]{#1\rbrak{#2}}							
\newcommand{\half}{\tfrac{1}{2}}							
\newcommand{\dd}{\mathrm{d}} 								
\newcommand{\dInt}[1]{{\dd {#1}}}		 					
\newcommand{\mindex}[1]{_\mathrm{#1}}						
\newcommand{\transpose}{^\intercal}							
\newcommand{\inverse}{^{-1}}								
\newcommand{\pdxy}[2]{\frac{\partial{#1}}{\partial{#2}}}	
\newcommand{\density}{\varrho}								
\newcommand{\velocity}{u}									
\newcommand{\dynvisc}{\eta}									
\newcommand{\rateofshear}{\dot{\gamma}}						

\newcommand{\tens}[1]{\mathsf{#1}}

\newcommand{\trace}[1]{\mathrm{tr}\,#1}
\newcommand{\symbolUndeformedCoordinate}{x}
\newcommand{\symbolDeformedCoordinate}{y}
\newcommand{\symbolDeformationGradient}{F}
\newcommand{\nodeindex}{\alpha}
\newcommand{\tetcoord}{\xi}
\newcommand{\symbolRightCGDT}{C}

\newcommand{\jacobian}{\tens{J}}
\newcommand{\undeformed}[1]{\symbolUndeformedCoordinate_{#1}}
\newcommand{\deformed}[1]{\symbolDeformedCoordinate_{#1}}

\newcommand{\defgrad}{\tens{\symbolDeformationGradient}}
\newcommand{\defgradij}[1]{\defgrad_{#1}}

\newcommand{\rightcauchy}{\tens{\symbolRightCGDT}}

\newcommand{\rightcauchyij}[1]{\rightcauchy_{#1}}

\newcommand{\displacement}{u}

\newcommand{\youngsmodulus}{E}
\newcommand{\shearmodulus}{\mu}
\newcommand{\bulkmodulus}{\kappa}
\newcommand{\poissonratio}{\nu}
\newcommand{\cellradius}{R}

\newcommand{\symbolShapeFunction}{N}
\newcommand{\shapefunc}[1]{\func{\symbolShapeFunction^{#1}}{\tetcoord_1,\tetcoord_2,\tetcoord_3}}
\newcommand{\shapefunca}{\func{\symbolShapeFunction^{\nodeindex}}{\tetcoord_1,\tetcoord_2,\tetcoord_3}}
\newcommand{\nhinvariantI}{I}
\newcommand{\nhinvariantJ}{J}
\newcommand{\nhinvariantK}{K}
\newcommand{\symbolForce}{f}
\newcommand{\force}{\symbolForce}

\newcommand{\strainenergydensity}{U}
\newcommand{\capillarynumber}{\mathrm{Ca}}

\newcommand{\taylordeformation}{D}
\newcommand{\volumefraction}{\phi}
\newcommand{\inertiatensor}{\Theta}
\newcommand{\mooneyrivlinratio}{w}
\newcommand{\strain}{\Delta\epsilon}
\newcommand{\new}[1]{#1}
\usepackage{hyperref}
\hypersetup{
	pdftitle={A hyperelastic model for simulating cells in flow},    
	pdfauthor={Sebastian J. M\"uller, Franziska Weigl, Carina Bezold, Ana Sancho, Christian B\"acher, Krystyna Albrecht, Stephan Gekle},     
	colorlinks=true,		
	linkcolor=blue,			
	citecolor=blue,			
	filecolor=magenta,		
	urlcolor=cyan			
}
\usepackage{tikz}
\usetikzlibrary{calc, arrows, decorations.markings}
\newcommand{\sisref}[1]{section~S-#1}
\newcommand{\sifref}[1]{figure~S-#1}
\begin{document}
	\thispagestyle{empty}
	\twocolumn[%
	\begin{@twocolumnfalse}
	\textbf{Published 2020 in: Biomechanics and Modeling in Mechanobiology}
	\\
	\textbf{DOI: }\href{https://doi.org/10.1007/s10237-020-01397-2}{\textbf{10.1007/s10237-020-01397-2}}
	\\
	\textit{Unfortunately, the journal version misses an 		important contributor.}
	\\ \textit{The correct author list is the one given in this document.}
	\vspace*{2em}
		\begin{center}
			{\LARGE A hyperelastic model for simulating cells in flow}
			\\ \vspace*{0.05\linewidth}
			\begin{minipage}{0.8\linewidth}
				{\large Sebastian J. M\"uller\,$^1$, Franziska Weigl\,$^2$, Carina Bezold\,$^1$, Ana Sancho\,$^{2,3}$, Christian B\"acher\,$^1$, Krystyna Albrecht\,$^2$ and Stephan Gekle\,$^1$}
			\end{minipage}
			\\ \vspace*{0.05\linewidth}
			\begin{minipage}{0.8\linewidth}
				$^1$ Theoretical Physics VI, Biofluid Simulation and Modeling, University of Bayreuth, Universit\"atsstra{\ss}e 30, 95440 Bayreuth, Germany
				\\
				$^2$ Department of Functional Materials in Medicine and Dentistry and Bavarian Polymer Institute (BPI), University of W\"urzburg, Pleicherwall 2, 97070 W\"urzburg, Germany
				\\
				$^3$ Department of Automatic Control and Systems Engineering, University of the Basque Country UPV/EHU, San Sebastian, Spain
			\end{minipage}
		\end{center}
		\begin{abstract}
			\noindent
			In the emerging field of 3D bioprinting, cell damage due to large deformations is considered a main cause for cell death and loss of functionality inside the printed construct. 
			Those deformations, in turn, strongly depend on the mechano-elastic response of the cell to the hydrodynamic stresses experienced during printing. 
			In this work, we present a numerical model to simulate the deformation of biological cells in arbitrary three-dimensional flows. 
			We consider cells as an elastic continuum according to the hyperelastic Mooney--Rivlin model. 
			We then employ force calculations on a tetrahedralized volume mesh. 
			
			To calibrate our model, we perform a series of FluidFM\textsuperscript{\textregistered}\  compression experiments with REF52 cells \new{demonstrating that all three parameters of the Mooney--Rivlin model are required for a good description of the experimental data at very large deformations up to $\SI{80}{\percent}$}.
			In addition, we validate the model by comparing to previous AFM experiments on bovine endothelial cells and artificial hydrogel particles. 
			To investigate cell deformation in flow, we incorporate our model into Lattice Boltzmann simulations via an Immersed-Boundary algorithm.
			In linear shear flows, our model shows excellent agreement with analytical calculations and previous simulation data.
		\end{abstract}
		\begin{minipage}{\linewidth}
		\end{minipage}
		\\ \vspace*{0.05\linewidth}
	\end{@twocolumnfalse}
	]
\section{Introduction}
\label{intro}
The dynamic behavior of flowing cells is central to the functioning of organisms and forms the base for a variety of biomedical applications.
Technological systems that make use of the elastic behavior of cells are, for example, cell sorting \citep{shen_recent_2019}, real-time deformability cytometry \citep{otto_real-time_2015,fregin_high-throughput_2019} or probing techniques for cytoskeletal mechanics \citep{kollmannsberger_linear_2011, GonzalezCruz_2012, Huber_2013, Bongiorno_2014, fischer-friedrich_quantification_2014, Lange_2015, fischer-friedrich_rheology_2016, Nyberg_2017, Lange_2017, Kubitschke_2017, jaiswal_stiffness_2017, Mulla_2019}.
In most, but not all, of these applications cell deformations typically remain rather small. 
A specific example where large deformations become important is 3D bioprinting.
Bioprinting is a technology which, analogously to common 3D printing, pushes a suspension of cells in highly viscous hydrogels---a so-called bioink---through a fine nozzle to create three-dimensional tissue structures.	
A major challenge in this process lies in the control of large cell deformations and cell damage during printing.
Those deformations arise from hydrodynamic stresses in the printer nozzle and ultimately affect the viability and functionality of the cells in the printed construct \citep{snyder_mesenchymal_2015,blaeser_controlling_2015,zhao_influence_2015,paxton_proposal_2017,muller_flow_2020-1}.
How exactly these hydrodynamic forces correlate with cell deformation, however, strongly depends on the elastic behavior of the cell and its interaction with the flowing liquid.
Theoretical and computational modeling efforts in this area have thus far been restricted to pure fluid simulations without actually incorporating the cells \citep{Khalil_2007, Aguado_2012, blaeser_controlling_2015} or simple 2D geometries \citep{Tirella_2011, Li_2015_bioprinting}.
The complexity of cell mechanics and the diversity of possible applications make theoretical modeling of cell mechanics in flow a challenge which, to start with, requires reliable experimental data for large cell deformations.

The most appropriate tool to measure cellular response at large deformations is atomic force microscopy (AFM) \citep{lulevich_deformation_2003, lulevich_cell_2006, ladjal_atomic_2009, kiss_elasticity_2011, fischer-friedrich_quantification_2014, Hecht_2015, Ghaemi_2016, Sancho_2017, Efremov_2017, Ladjal_2018, Chim_2018}.
AFM cantilevers with pyramidal tips, colloidal probes, or flat geometries are used to indent or compress cells.
Therefore, a common approach to characterize the elasticity of cells utilizes the Hertzian theory, which describes the contact between two linear elastic solids \citep[p.~90-104]{johnson_contact_2003}, but is limited to the range of small deformations \citep{dintwa_accuracy_2008}. 
Experimental measurements with medium-to-large deformations typically show significant deviations from the Hertz prediction, \eg for cells or hydrogel particles \citep{ neubauer_mechanoresponsive_2019-1}.
Instead of linear elasticity, a suitable description of cell mechanics for bioprinting applications requires more advanced hyperelastic material properties.
While for simple anucleate fluid-filled cells such as, e.g., red blood cells, theoretical models abound \citep{Freund_2014, Zavodszky_2017, Mauer_2018, Guckenberger_2018_phase, Kotsalos_2019}, the availability of models for cells including a complex cytoskeleton is rather limited.
In axisymmetric geometries, \citet{caille_contribution_2002-1} and \citet{Mokbel_2017} used an axisymmetric finite element model with neo-Hookean hyperelasticity to model AFM and microchannel experiments on biological cells.
In shear flow, \new{approximate analytical treatments are possible \citep{Roscoe_1967, gao_deformation_2009, Gao_2011_JFM, Gao_2012_PRL}}. 
Computationally, \citet{gao_deformation_2009} carried out 2D simulations while in 3D \citet{Lykov_2017} utilized a DPD technique based on a bead-spring model.  Furthermore, \citet{Villone_2014_compFluid, Villone_2015} presented an arbitrary Lagrangian-Eulerian approach for elastic particles in viscoelastic fluids. 
Finally, \citet{rosti_rheology_2018} and \citet{saadat_immersed-finite-element_2018} considered viscoelastic and neo-Hookean finite element models, respectively, in shear flow.


In this work, we introduce and \new{calibrate} a computational model for fully three-dimensional simulations of cells in arbitrary flows. 
\new{Our approach uses a Lattice-Boltzmann solver for the fluid and a direct force formulation for the elastic equations.
In contrast to earlier works \citep{caille_contribution_2002-1, Gao_2011_JFM, Mokbel_2017, rosti_rheology_2018, saadat_immersed-finite-element_2018} our model uses a three-parameter Mooney--Rivlin elastic energy functional.
To demonstrate the need for this more complex elastic model, we carry out extensive FluidFM\textsuperscript{\textregistered}\  indentation experiments for REF52 (rat embryonic fibroblast) cells at large cell deformation up to $\SI{80}{\percent}$ \citep{alexandrova_comparative_2008}.
}
In addition, our model compares favorably with previous AFM experiments on bovine endothelial cells \citep{caille_contribution_2002-1} as well as artificial hydrogel particles \citep{neubauer_mechanoresponsive_2019-1}.
Our model provides a much more realistic force--deformation behavior compared to the small-deformation Hertz approximation, but is still simple and fast enough to allow the simulation of dense cell suspensions in reasonable time.
Particularly, our approach is less computationally demanding than conventional finite-element methods which usually require large matrix operations.	
Furthermore, it is easily extensible and allows, e.g., the inclusion of a cell nucleus by the choice of different elastic moduli for different parts of the volume.

We finally present simulations of our cell model in different flow scenarios using an Immersed-Boundary algorithm to couple our model with Lattice Boltzmann fluid calculations.
In a plane Couette (linear shear) flow, we investigate the shear stress dependency of single cell deformation, which we compare to the average cell deformation in suspensions with higher volume fractions, and show that our results in the neo-Hookean limit are in accordance with earlier elastic cell models \citep{Gao_2011_JFM, rosti_rheology_2018, saadat_immersed-finite-element_2018}.
\section{Theory}
\label{sec:theory}
In general, hyperelastic models are used to describe materials that respond elastically to large deformations \citep[p.~93]{bower_applied_2010}.
Many cell types can be subjected to large reversible shape changes.
This section provides a brief overview of the hyperelastic Mooney--Rivlin model implemented in this work.

The displacement of a point is given by
\begin{align}
\label{eq:displacement-field}
\displacement_i = \deformed{i} - \undeformed{i} \eqspace ,
\end{align}
where $\undeformed{i}$ ($i=1,2,3$) refers to the undeformed configuration (material frame) and $\deformed{i}$ to the deformed coordinates (spatial frame).
We define the deformation gradient tensor and its inverse as \citep[p.~14,18]{bower_applied_2010}
\begin{align}
\label{eq:displacement-gradient-tensor}
\defgradij{ij} = \pdxy{\deformed{i}}{\undeformed{j}} = \pdxy{\displacement_i}{\undeformed{j}} + \delta_{ij}
\quad \mathrm{and} \quad \defgradij{ij}\inverse = \pdxy{\undeformed{i}}{\deformed{j}}\eqspace .
\end{align}
Together with the right Cauchy-Green deformation tensor, $\rightcauchy = \defgrad\transpose\defgrad$ (material description), we can define the following invariants which are needed for the strain energy density calculation below:
\begin{align}
\nhinvariantJ & = \det \defgrad  \label{eq:invariantJ} \\
\nhinvariantI & = T_{\rightcauchy} \nhinvariantJ^{-2/3} \label{eq:invariantI} \\
\nhinvariantK & = \half \rbrak{ T_{\rightcauchy}^2 - T_{\rightcauchy^2}} \nhinvariantJ^{-4/3} \label{eq:invariantK}  
\end{align}
Here,
\begin{align}
T_{\rightcauchy} = \trace{\rightcauchy} \quad \mathrm{and} \quad T_{\rightcauchy^2} = \trace{\rbrak{\rightcauchy^2}}
\end{align}
are the trace of the right Cauchy-Green deformation tensor and its square, respectively.
The nonlinear strain energy density of the Mooney--Rivlin model is given by \citep{mooney_theory_1940,rivlin_large_1948}
\begin{align}
\label{eq:strain-energy-density:mr}
\strainenergydensity =\sbrak{ \frac{\shearmodulus_1}{2} \rbrak{\nhinvariantI - 3} + \frac{\shearmodulus_2}{2} \rbrak{\nhinvariantK - 3} + \frac{\bulkmodulus}{2} \rbrak{\nhinvariantJ-1}^2} \eqspace ,
\end{align}
where $\shearmodulus_1$, $\shearmodulus_2$, and $\kappa$ are material properties.
They correspond---for consistency with linear elasticity in the range of small deformations---to the shear modulus $\shearmodulus=\shearmodulus_1+\shearmodulus_2$ and bulk modulus $\kappa$ of the material and are therefore related to the Young's modulus $\youngsmodulus$ and the Poisson ratio $\poissonratio$ via \citep[p.~74]{bower_applied_2010}
\begin{align}
\label{eq:elastic-moduli:mr}
\mu = \frac{\youngsmodulus}{2\rbrak{1+\poissonratio}} \quad \mathrm{and} \quad
\kappa = \frac{\youngsmodulus}{3\rbrak{1-2\poissonratio}} \eqspace .
\end{align}
Through the choice $\shearmodulus_2=0$ in \eqref{eq:strain-energy-density:mr}, we recover the simpler and frequently used \citep{Gao_2011_JFM,saadat_immersed-finite-element_2018} neo-Hookean strain energy density:
\begin{align}
\label{eq:strain-energy-density:nh}
\strainenergydensity\mindex{NH} =\sbrak{ \frac{\shearmodulus}{2} \rbrak{\nhinvariantI - 3} + \frac{\bulkmodulus}{2} \rbrak{\nhinvariantJ-1}^2}
\end{align}
As we show later, this can be a sufficient description for some cell types.
To control the strength of the second term and quickly switch between neo-Hookean and Mooney--Rivlin strain energy density calculation, we introduce a factor $\mooneyrivlinratio \in \sbrak{0,1}$ and set
\begin{align}
\shearmodulus_1 = \mooneyrivlinratio \shearmodulus \quad \mathrm{and} \quad \shearmodulus_2 = (1-\mooneyrivlinratio)\shearmodulus \eqspace 
\end{align}
such that $\mooneyrivlinratio=1$, which equals setting $\shearmodulus_2=0$ in \eqref{eq:strain-energy-density:mr}, corresponds to the purely neo-Hookean description in \eqref{eq:strain-energy-density:nh}, while $\mooneyrivlinratio<1$ increases the influence of the $\shearmodulus_2$-term \new{and thus leads to a more pronounced strain hardening as shown in \sifref{6} of the Supporting Information.}
%
\section{Tetrahedralized cell model}
\label{sec:FEmodel}
In this section we apply the hyperelastic theory of \sref{sec:theory} to a tetrahedralized mesh as shown in \fref{fig:cell-inner-grid}.
\subsection{Calculation of elastic forces}
We consider a mesh consisting of tetrahedral elements as depicted in \fref{fig:cell-inner-grid}.
The superscript $\nodeindex$ refers to the four vertices of the tetrahedron.
The elastic force acting on vertex $\nodeindex$ in direction $i$ is obtained from \eqref{eq:strain-energy-density:mr} by differentiating the strain energy density $\strainenergydensity$ with respect to the vertex displacement as
\begin{align}
\label{eq:elastic-force-vertex-0}
\force_i^\nodeindex  = & - V_0 \pdxy{\strainenergydensity}{\displacement_i^\nodeindex} \eqspace ,
\end{align}
where $V_0$ is the reference volume of the tetrahedron.
In contrast to \citet{saadat_immersed-finite-element_2018}, the numerical calculation of the force in our model does not rely on the integration of the stress tensor, but on a differentiation where the calculation of all resulting terms involves only simple arithmetics.
Applying the chain rule for differentiation yields:
\begin{align}
\label{eq:elastic-force-vertex-1}
\force_i^\nodeindex  = -V_0 
\Bigg[ & 
\rbrak{ \pdxy{\strainenergydensity}{\nhinvariantI} \pdxy{\nhinvariantI}{T_{\rightcauchy}} + \pdxy{\strainenergydensity}{\nhinvariantK} \pdxy{\nhinvariantK}{T_{\rightcauchy}}} \pdxy{T_{\rightcauchy}}{\defgradij{kl}}
\nonumber \\ &
+\rbrak{ \pdxy{\strainenergydensity}{\nhinvariantI} \pdxy{\nhinvariantI}{\nhinvariantJ} + \pdxy{\strainenergydensity}{\nhinvariantK} \pdxy{\nhinvariantK}{\nhinvariantJ} + \pdxy{\strainenergydensity}{\nhinvariantJ} } \pdxy{\nhinvariantJ}{\defgradij{kl}}
\nonumber \\ &
+ \pdxy{\strainenergydensity}{\nhinvariantK} \pdxy{\nhinvariantK}{T_{\rightcauchy^2}} \pdxy{T_{\rightcauchy^2}}{\defgradij{kl}}
\Bigg] \pdxy{\defgradij{kl}}{\displacement_i^\nodeindex}
\end{align}
The evaluation of \eqref{eq:elastic-force-vertex-1} requires the calculation of the deformation gradient tensor $\defgrad$, which is achieved by linear interpolation of the coordinates and displacements inside each tetrahedral mesh element as detailed in the next section.
\new{We note that our elastic force calculation is purely local making it straightforward to employ different elastic models in different regions of the cell and/or to combine it with elastic shell models.
This flexibility can be used to describe, e.g., the cell nucleus \citep{caille_contribution_2002-1} or an actin cortex \citep{Bacher_2019} surrounding the cell interior.}
\subsection{Interpolation of the displacement field}

\begin{figure*}
	\centering
	(a)
	\includegraphics[width=0.25\linewidth]{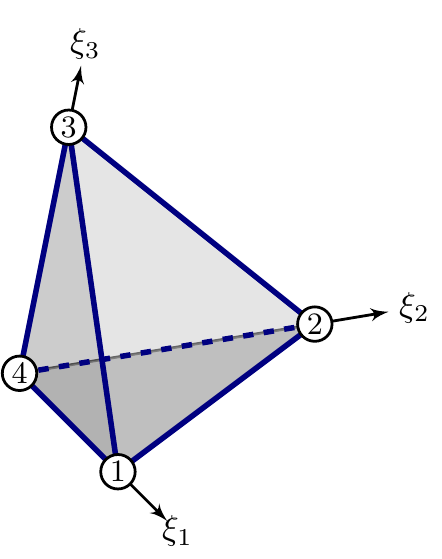}
	\qquad
	(b)
	\includegraphics[width=0.25\linewidth]{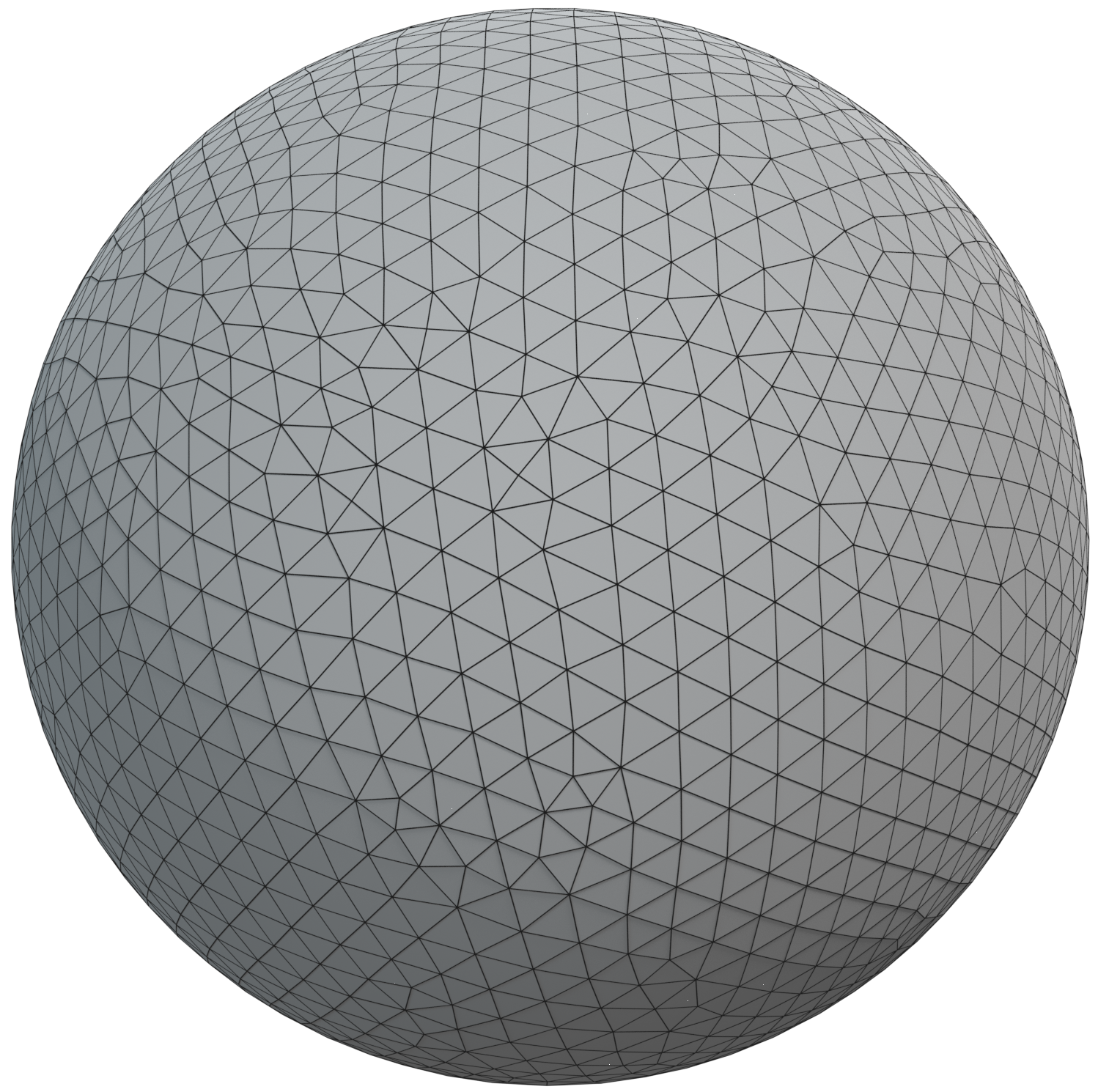}		
	\qquad
	(c)
	\includegraphics[width=0.25\linewidth]{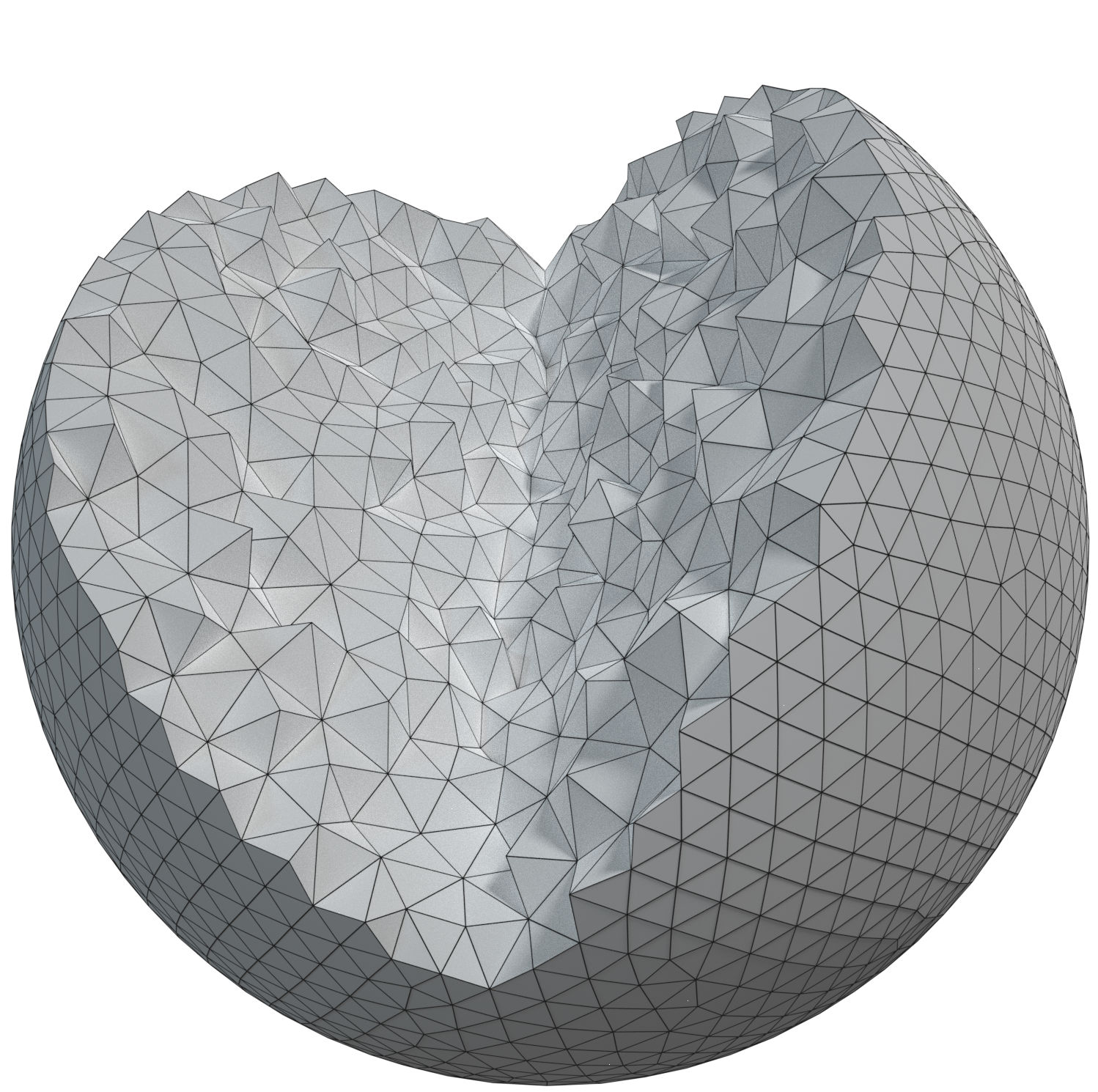}
	\caption{(a) The four noded tetrahedron as mesh element within a local dimensionless coordinate system $\cbrak{\tetcoord_1,\tetcoord_2,\tetcoord_3}$. (b) The spherical cell model with its triangulated surface and (c) its inner tetrahedralized mesh
	}
\label{fig:cell-inner-grid}	
\end{figure*}

Following standard methods, e.g. \citet{bower_applied_2010}, we start by interpolating a point $\undeformed{i}$ inside a single tetrahedron using the vertex positions $\undeformed{i}^\nodeindex$ ($\nodeindex=1,2,3,4$).
The interpolation uses an inscribed, dimensionless coordinate system, denoted by $\rbrak{\tetcoord_1,\tetcoord_2,\tetcoord_3}$ with $0\leq\tetcoord_i\leq1$\footnote{\citet[p.~481,483]{bower_applied_2010} erroneously states a range of $-1\leq\tetcoord_i\leq1$ for the tetrahedral element.}, as depicted in \fref{fig:cell-inner-grid}a.
One vertex defines the origin while the remaining three indicate the coordinate axes.
A set of shape functions, \ie interpolation functions, $\shapefunca$ is employed to interpolate positions inside the tetrahedron volume.
An arbitrary point $\undeformed{i}$ inside the element is interpolated as
\begin{align}
\label{eq:interpolation-point}
\undeformed{i} = \sum\limits_{\nodeindex=1}^{4} \shapefunca \undeformed{i}^\nodeindex \eqspace ,
\end{align}
where the shape functions are defined as \citep[p.~483]{bower_applied_2010}:
\begin{align}
\shapefunc{1} & = \tetcoord_1 \\
\shapefunc{2} & = \tetcoord_2 \\
\shapefunc{3} & = \tetcoord_3 \\
\shapefunc{4} & = 1-\tetcoord_1-\tetcoord_2-\tetcoord_3
\end{align}
According to \eqref{eq:displacement-field}, the displacement of vertex $\nodeindex$ in $i$-direction is given by
\begin{align}
\label{eq:vertex-displacement}
\displacement_i^\nodeindex = \deformed{i}^\nodeindex - \undeformed{i}^\nodeindex \eqspace .
\end{align}
Therefore similar to \eqref{eq:interpolation-point}, the displacement at an arbitrary point in the volume can also be expressed in terms of the shape functions and the vertex displacements as
\begin{align}
\label{eq:interpolation-displacement}
\displacement_i = \sum\limits_{\nodeindex=1}^{4} \shapefunca \displacement_i^\nodeindex \eqspace .
\end{align}
The calculation of the deformation gradient tensor according to \eqref{eq:displacement-gradient-tensor} requires the spatial derivative of the displacement:
\begin{align}
\defgradij{ij} - \delta_{ij} = \pdxy{\displacement_i}{\undeformed{j}}  
\label{eq:displacement-derivative}
 & = \pdxy{\displacement_i}{\tetcoord_k}\pdxy{\tetcoord_k}{\undeformed{j}} = \tens{A}_{ik} \tens{B}_{kj} 
\end{align}
By inserting \eqref{eq:interpolation-displacement} into \eqref{eq:displacement-derivative} and evaluating the shape functions, the components of the matrix $\tens{A}$ are easily determined to be the difference of the displacements between the origin (vertex 4) and the remaining vertices 1, 2 and 3:
\begin{align}
\label{eq:matrix-A}
\tens{A}_{ik} = \displacement_i^k - \displacement_i^4
\end{align}
Note that due to the linear interpolation $\tens{A}_{ik}$ is constant inside a given tetrahedron.
The matrix $\tens{B}=\tens{\jacobian}\inverse$ is the inverse of the Jacobian matrix, obtained similarly to \eqref{eq:matrix-A} as
\begin{align}
\jacobian_{ik} & = \pdxy{\undeformed{i}}{\tetcoord_k} = \undeformed{i}^k - \undeformed{i}^4 \eqspace .
\end{align}
%
Since $\undeformed{i}$ refers to the reference coordinates, the calculation of the matrices $\tens{\jacobian}$ and $\tens{B}$ has to be performed only once at the beginning of a simulation.
With the interpolation of the displacement in each tetrahedron, we can write all derivatives occurring in \eqref{eq:elastic-force-vertex-1}, as listed in the following: 
{\\ \renewcommand{\arraystretch}{2}
	\begin{tabular}{l l}
		$\pdxy{\strainenergydensity}{\nhinvariantI} = \frac{\shearmodulus_1}{2} $ &
		$\pdxy{\nhinvariantI}{T_{\rightcauchy}}                  = \nhinvariantJ^{-\frac{2}{3}}$ \\
		$\pdxy{\strainenergydensity}{\nhinvariantK} = \frac{\shearmodulus_2}{2} $ &
		$\pdxy{\nhinvariantK}{T_{\rightcauchy}}       = T_{\rightcauchy} \nhinvariantJ^{-\frac{4}{3}}$ \\
		$\pdxy{T_{\rightcauchy}}{\defgradij{il}}                 = 2 \defgradij{il}$ &
		$\pdxy{\nhinvariantI}{\nhinvariantJ}        = -\frac{2}{3} T_{\rightcauchy} \nhinvariantJ^{-\frac{5}{3}}$ \\
		$\pdxy{\nhinvariantK}{\nhinvariantJ}	    = -\frac{2}{3} \rbrak{T_{\rightcauchy}^2 - T_{\rightcauchy^2}} \nhinvariantJ^{-\frac{7}{3}}$ &
		$\pdxy{\strainenergydensity}{\nhinvariantJ} = \kappa \rbrak{\nhinvariantJ-1}$ \\
		$\pdxy{\nhinvariantJ}{\defgradij{il}}       =  \nhinvariantJ \defgradij{li}\inverse$ &
		$\pdxy{\nhinvariantK}{T_{\rightcauchy^2}}       = -\half \nhinvariantJ^{-\frac{4}{3}}$ \\
		$\pdxy{T_{\rightcauchy^2}}{\defgradij{il}}             =  4 \defgradij{ik} \rightcauchyij{kl}$ &
		$\pdxy{\defgradij{kl}}{\displacement_i^\nodeindex} = \delta_{ki} \tens{B}_{ml} \rbrak{ \delta_{m \nodeindex} - \delta_{4\nodeindex} }$ \\
	\end{tabular}
}
\subsection{Taylor deformation parameter}
As a measure for the cell deformation, we use the Taylor deformation parameter \citep{ramanujan_deformation_1998,clausen_capsule_2010, Guckenberger_2016, saadat_immersed-finite-element_2018} 
\begin{align}
\label{eq:taylor-deformation-parameter-definition}
\taylordeformation = \frac{a_3 - a_1}{a_3 + a_1} \eqspace ,
\end{align}
where $a_1$ and $a_3$ are respectively the minor and major semi axis of an ellipsoid corresponding to the inertia tensor of the cell.
The Taylor deformation is a good measure for approximately elliptic cell deformations, as they occur in shear flow (cf.~\sref{sec:shear-flow-simulations}).

To calculate $\taylordeformation$, first the components of the inertia tensor
\begin{align}
\label{eq:inertia-tensor-integral}
\tens{\inertiatensor}_{ij} = \int\limits_V  x_k x_k \delta_{ij} - x_i x_j \dInt{V} \eqspace ,
\end{align}
where $\vec{x}$ is a vector inside the volume $V$, are calculated using our discretized cell with $N\mindex{tet}$ tetrahedra as
\begin{align}
\label{eq:inertia-tensor-discretized}
\tens{\inertiatensor}_{ij} = \sum\limits_{l=1}^{N\mindex{tet}} V_l \rbrak{r_k^l r_k^l \delta_{ij} - r_i^l r_j^l}  \eqspace .
\end{align}
The vector $\vec{r}^{\,l}$ denotes the center of mass of the $l^\mathrm{th}$ tetrahedron and $V_l$ is its current volume.
The eigenvalues $\theta_1 > \theta_2 > \theta_3$ of $\tens{\inertiatensor}$ can be used to fit the semi axes $a_1 < a_2 < a_3$ of the corresponding ellipsoid:
\begin{align}
\label{eq:inertia-tensor-ellipsoid-semiaxes}
a_1 & = \frac{5}{2 M} \rbrak{- \theta_1 + \theta_2 + \theta_3} \nonumber \\
a_2 & = \frac{5}{2 M} \rbrak{  \theta_1 - \theta_2 + \theta_3} \nonumber \\
a_3 & = \frac{5}{2 M} \rbrak{  \theta_1 + \theta_2 - \theta_3}
\end{align}
The prefactor contains the mass $M$ of the ellipsoid (considering uniform mass density) and drops out in the calculation of $\taylordeformation$.
%
%
\section[Comparison to FluidFM\textsuperscript{\textregistered}\ measurements]{Comparison of the numerical model to FluidFM\textsuperscript{\textregistered}\  measurements on REF52 cells}
\label{sec:fluidfm-section}
In this section, we validate compression simulations of our cell model with FluidFM\textsuperscript{\textregistered}\  compression experiments of REF52 cells stably expressing paxillin-YFP \citep{alexandrova_comparative_2008}.
These experiments provide as an output the required force to produce a certain deformation of the cell, which can be directly compared to our model.
We start with a detailed description of the experiments and show the suitability of our model to describe the elastic behavior of REF52 cells afterwards.
\subsection{FluidFM\textsuperscript{\textregistered}\  indentation measurements}
\label{sec:fluidfm-measurements}	
We perform a series of compression measurements of REF52 cells with a Flex FPM (Nanosurf GmbH, Germany) system that combines the AFM with the FluidFM\textsuperscript{\textregistered}\  technology (Cytosurge AG, Switzerland).
\new{In contrast to conventional AFM techniques, FluidFM\textsuperscript{\textregistered}\  uses flat cantilevers that possess a microchannel connected to a pressure system.}
By applying a suction pressure, cells can be aspirated and retained at the aperture of the cantilever’s tip. A more detailed description of the setup and its functionality is already reported in \cite{Sancho_2017}.
\new{All experiments are based on a cantilever with an aperture of $\SI{8}{\micro\meter}$ diameter and a nominal spring constant of $\SI{2}{\newton\per\meter}$. In order to measure the cellular deformation, a cell was sucked onto the tip and compressed between the cantilever and the substrate until a setpoint of $\SI{100}{\nano\newton}$ was reached. Immediately before the experiment, the cells were detached by using Accutase (Sigma Aldrich) and were therefore in suspension at the time of indentation. In this way, it can be ensured that only a single cell is deformed during each measurement.}

\begin{figure}
	\centering
	\includegraphics[width=0.75\linewidth]{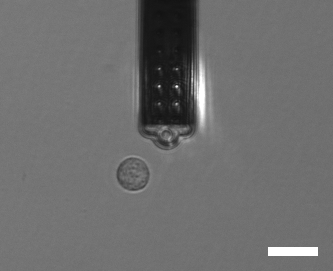}
	\caption{Example micrograph showing the FluidFM\textsuperscript{\textregistered}\  cantilever and a cell viewed from the top. Scale bar is $\SI{30}{\micro\meter}$}
	\label{fig:fluidfm}	
\end{figure}

An example micrograph of the experiment before compression is shown in \fref{fig:fluidfm}.
Analogously to AFM, primary data in form of cantilever position (in $\si{m}$) and deflection (in $\si{\volt}$) has to be converted to force and deformation through the deflection sensitivity (in $\si{\meter\per\volt}$) and the cantilevers' spring constant.
The cellular deformation further requires the determination of the contact point, which we choose as the cantilever position where the measured force starts to increase.
The undeformed cell size is obtained as mean from a horizontal and vertical diameter measurement using the software imageJ. 
\subsection{Simulation setup}
\label{sec:simulation-setup}

\begin{figure*}
	\centering
	(a)
	\includegraphics[width=0.2\linewidth]{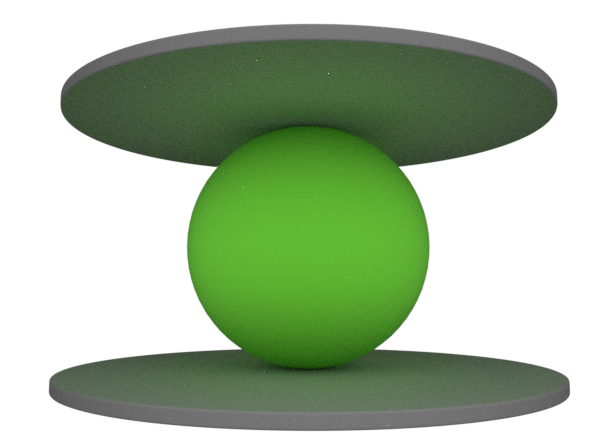}
	(b)
	\includegraphics[width=0.2\linewidth]{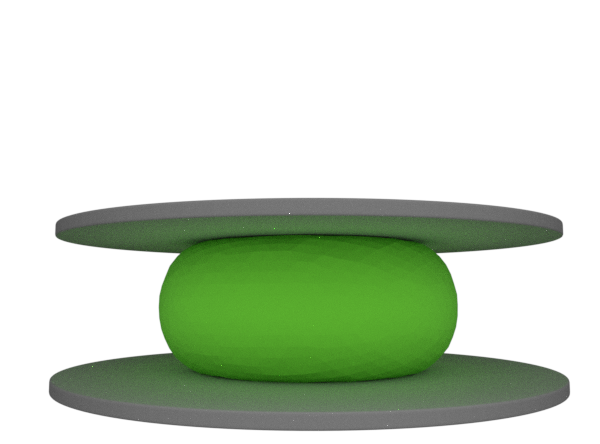}
	(c)
	\includegraphics[width=0.2\linewidth]{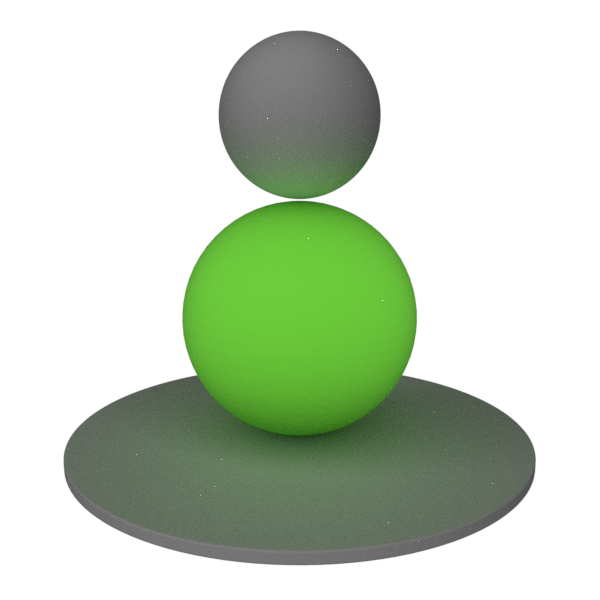}
	(d)
	\includegraphics[width=0.2\linewidth]{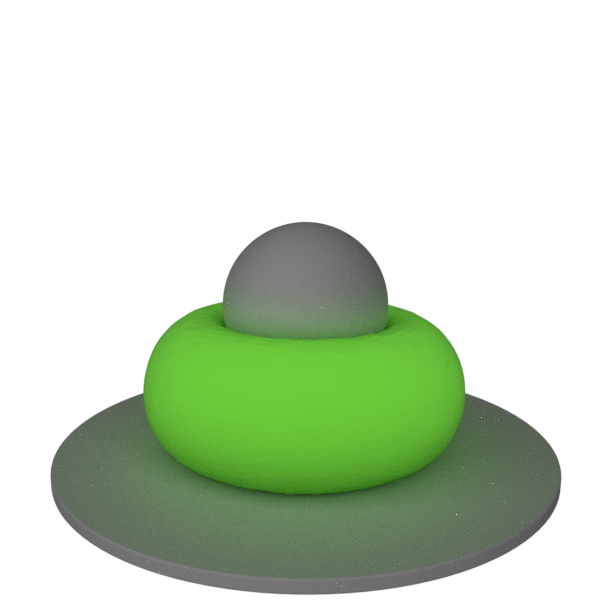}
	\caption{(a and b) Cell compression simulations: The cell is compressed between a lower, resting, and an upper, moving, plate. (c and d) Colloidal probe cell indentation simulations: The cell rests on a plate, while being indented with a sphere
	}
\label{fig:compress-simulations}	
\end{figure*}

The experimental setup of the previous section is easily transferred and implemented for our cell model:
the undeformed spherical cell rests on a fixed plate while a second plate approaches from above to compress the cell as depicted in \fref{fig:compress-simulations} (a and b). 
In section~\ref{sec:validationAFM} below we will also use a slightly modified version where a sphere indents the cell as shown in \fref{fig:compress-simulations}~(c and d). 
A repulsive force prevents the cell vertices from penetrating the plates or the spherical indenter.
The elastic restoring forces (cf.\ \sref{sec:FEmodel}) acting against this imposed compression are transmitted throughout the whole mesh, deforming the cell.

We use meshes consisting of $\num{2000}$ to $\num{5000}$ vertices and about $\num{10000}$ to $\num{30000}$ tetrahedra to build up a spherical structure.
More details of the mesh and its generation (\sisref{2.4}) as well as the algorithm (\sisref{3}) are provided in the SI.
\subsection{Results}
\begin{figure}
	\includegraphics[width=\linewidth]{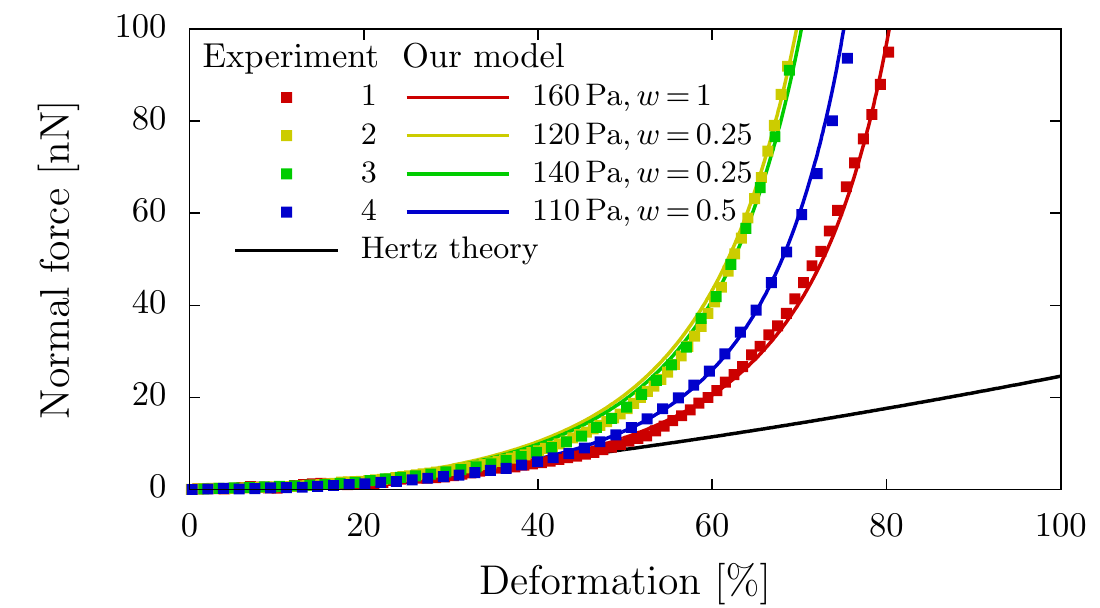}	
	\caption{Our numerical model in comparison to our FluidFM\textsuperscript{\textregistered}\  measurements on REF52 cells.
		The labels give the two fit parameters $\youngsmodulus$ and $\mooneyrivlinratio$.
		We find Young's moduli in the range of $\SI{110}{\pascal}$ to $\SI{160}{\pascal}$.
		The Hertz theory is shown for a Young's modulus of $\SI{180}{\pascal}$}
	\label{fig:comparison-fluidfm}
\end{figure}

In our FluidFM\textsuperscript{\textregistered}\  experiment series with REF52 cells, the cell radii lie between $\SI{7.1}{\micro\meter}$ and $\SI{10.4}{\micro\meter}$ with an overall average of $\SI{8.6\pm0.7}{\micro\meter}$.
In \fref{fig:comparison-fluidfm} we depict the force as function of the non-dimensionalized deformation, \ie the absolute compression divided by the cell diameter.
The experimental data curves share general characteristics: The force increases slowly in the range of small deformations up to roughly $\SI{40}{\percent}$, while a rapidly increasing force is observed for larger deformations.
Although the variation of the cell radius in the different measurements is already taken into account in the deformation, the point of the force upturn differs significantly which indicates a certain variability in the elastic parameters of the individual cells.

We use the compression simulation setup as detailed in \sref{sec:simulation-setup} to calculate force--deformation curves of our cell model.
The Poisson ratio is chosen as $\poissonratio=0.48$.
\new{In \sisref{2.7} of the Supporting Information we show that variations of the $ \poissonratio$ do not strongly affect the results.}
A best fit approach is used to determine the Young's modulus and the ratio of shear moduli $\mooneyrivlinratio$ and leads to very good agreement between model prediction and experimental data as shown in figure~\ref{fig:comparison-fluidfm} as well as \sisref{1} of the SI.
While the general range of force values is controlled using the Young's modulus, the Mooney--Rivlin ratio $\mooneyrivlinratio$ especially defines the point of the force upturn.
We find Young's moduli in the range $\SI{110}{\pascal}$ to $\SI{160}{\pascal}$ and $\mooneyrivlinratio=0.25$, $0.5$, and $1$.
For very small deformations our hyperelastic model produces the same results as would be expected from a linear elastic model according to the Hertz theory.
See the SI (\sisref{2.5}) for further details on the calculation of the force--deformation according to the Hertzian theory.
For large deformations, the force rapidly increases due to its nonlinear character, showing strain-hardening behavior and huge deviations from the Hertz theory.
Overall, we find an excellent match between simulation and our FluidFM\textsuperscript{\textregistered}\  measurements with REF52 cells.
\section[Comparison to other setups]{Comparison of our numerical model to other micromechanical setups}
In this section, we compare our simulations to axisymmetric calculations using the commercial software Abaqus and validate our cell model with further experimental data for bovine endothelial cells from \citep{caille_contribution_2002-1} and very recent data for hydrogel particles from \citep{neubauer_mechanoresponsive_2019-1}.
\subsection{Validation with axisymmetric simulations}
\label{sec:validation-abaqus}
To validate our model numerically, we compare our simulated force--deformation curves to calculations using the commercial software Abaqus \citep{abaqus_user_manual_2009} (version 6.14).
 
\begin{figure}
	\includegraphics[width=\linewidth]{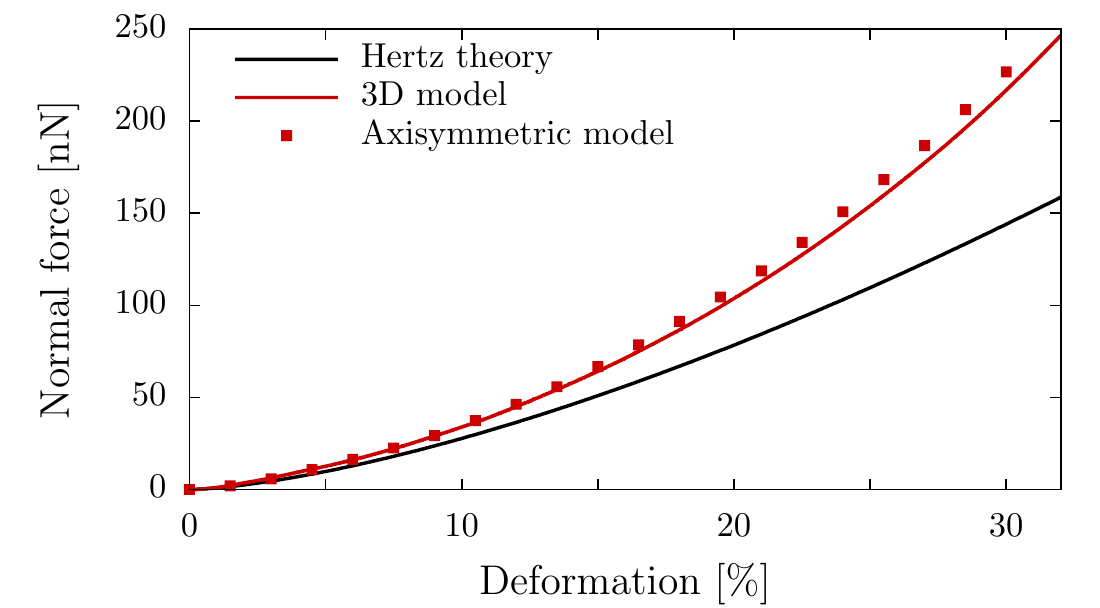}
	\caption{Comparison of force--deformation curves obtained from our model (red line) with the linear elastic Hertz theory (black line) and the two-dimensional simulation with Abaqus (red squares), showing good agreement between our three-dimensional and the axisymmetric model
}
\label{fig:validation-abaqus}
\end{figure}

In Abaqus, we use a rotationally symmetric setup consisting of a two-dimensional semicircle, which is compressed between two planes, similar to our simulation setup in \sref{sec:simulation-setup} and the finite element model utilized in \citep{caille_contribution_2002-1}.
The semicircle has a radius $r=\SI{15}{\micro\meter}$, a Young's modulus of $E=\SI{2.25}{\kilo\pascal}$ and a Poisson ratio of $\poissonratio=0.48$.
We choose a triangular mesh and the \mbox{built-in} implementation of the hyperelastic neo-Hookean model.
In \fref{fig:validation-abaqus} we see very good agreement between the results of the two different numerical methods.
\subsection{Validation with AFM experiments}
\label{sec:validationAFM}
To compare with the AFM experiments of \citet{caille_contribution_2002-1}, we simulate a cell with radius $\SI{15}{\micro\meter}$ using the setup of section~\ref{sec:simulation-setup}.
For the hydrogel particle indentation \citep{neubauer_mechanoresponsive_2019-1} we use the setup depicted in figure~\ref{fig:compress-simulations}~(c and d) with a particle radius of $\SI{40}{\micro\meter}$ and a radius of the colloidal probe of $\SI{26.5}{\micro\meter}$.
The Poisson ratio is chosen as $\num{0.48}$ in all simulations and the Young's modulus is determined using a best fit to the experimental data points.
Since the neo-Hookean description appears to be sufficient for these data sets, we further set $\mooneyrivlinratio=1$.

\begin{figure}
	(a)\\
	\includegraphics[width=\linewidth]{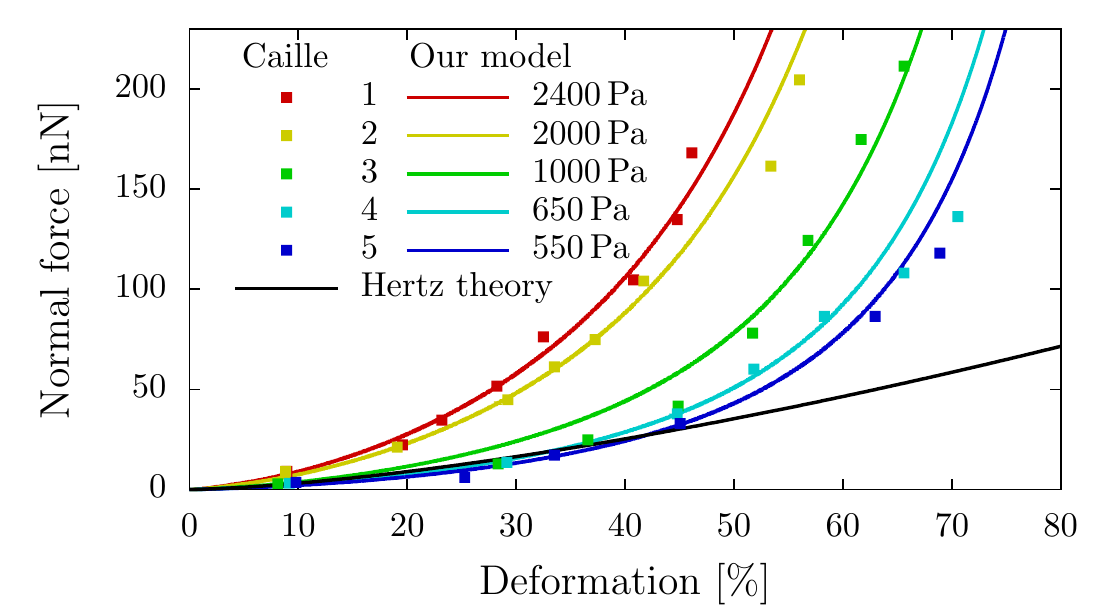}
	(b)\\
	\includegraphics[width=\linewidth]{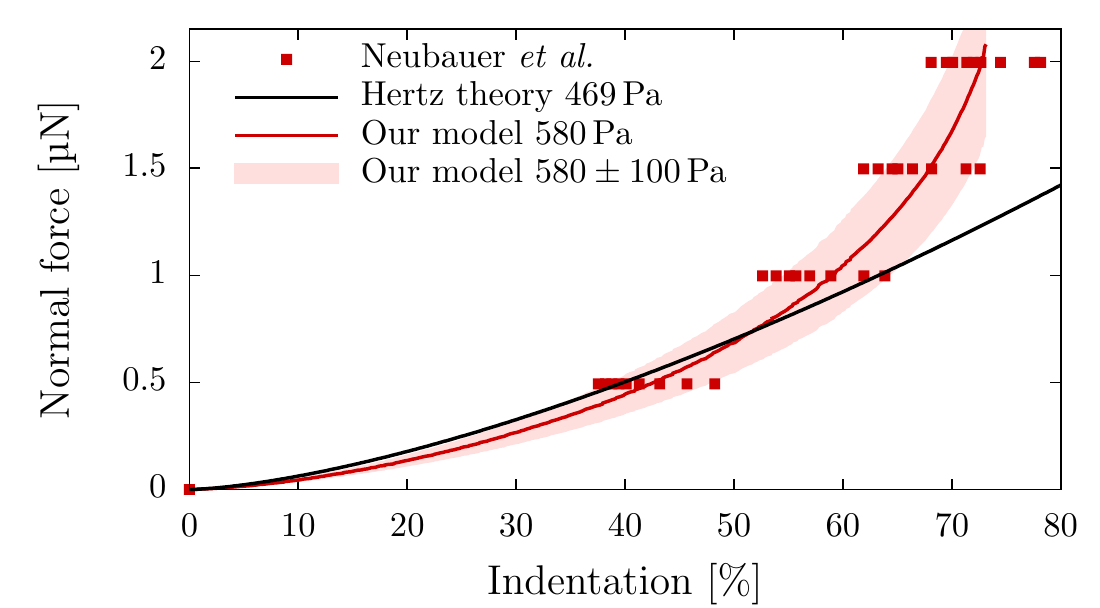}
	\caption{(a) Our numerical model in comparison to experimental measurements of bovine endothelial cells from \citep{caille_contribution_2002-1}. The black line depicts the prediction of the Hertz theory for a Young's modulus of $\SI{1000}{\pascal}$.
	(b) Our numerical model in comparison to experimental measurements of hydrogel particles from \citep{neubauer_mechanoresponsive_2019-1}. The indicated range corresponds to the experimentally found range of $\pm\SI{100}{\pascal}$ for the Young's modulus according to the depicted Hertz model
	}
	\label{fig:comparison-caille-neubauer}
\end{figure}

In \fref{fig:comparison-caille-neubauer}a, we show the experimental data for suspended, round, bovine endothelial cells of five separate measurements from \citep{caille_contribution_2002-1} together with the prediction of the Hertz theory for a Young's modulus of $\SI{1000}{\pascal}$. 
Fitting our data with Young's moduli in the range of $\SI{550}{\pascal}$ to $\SI{2400}{\pascal}$, we find good agreement between our calculations and the experimental data.
We note that \citet{caille_contribution_2002-1} observed similarly good agreement for their axisymmetric incompressible neo-Hookean FEM simulations which, however, cannot be coupled to external flows in contrast to the approach presented here. 
The same procedure is applied to the colloidal probe indentation data of hydrogel particles from \citep{neubauer_mechanoresponsive_2019-1}, showing in \fref{fig:comparison-caille-neubauer}b the experimental data and the prediction of the Hertz theory from \citep{neubauer_mechanoresponsive_2019-1}.
We find excellent agreement between our model calculations for Young's moduli in the range of $\num{580}\pm\SI{100}{\pascal}$ and the experimental data.
For both systems, \fref{fig:comparison-caille-neubauer} shows large deviations between the Hertzian theory and the experimental data for medium-to-large deformations. 
Our model provides a significant improvement in this range.
\section{Application in shear flow}
\label{sec:shear-flow-simulations}
We now apply our model to study the behavior of cells in a plane Couette (linear shear) flow setup and compare the steady cell deformation to other numerical and analytical cell models of \citet{Gao_2011_JFM}, \citet{rosti_rheology_2018} and \citet{saadat_immersed-finite-element_2018}.
A sketch of the simulation setup is shown in \fref{fig:singlecell-shear-simulation}.
For simplicity, we choose $\mooneyrivlinratio=1$ to reduce the Mooney--Rivlin description \eqref{eq:strain-energy-density:mr} to two free parameters $\shearmodulus$ and $\bulkmodulus$ (or $\youngsmodulus$ and $\poissonratio$),  obtaining a compressible neo-Hookean form.
We use the Lattice Boltzmann implementation of the open source software package ESPResSo \citep{limbach_espressoextensible_2006,roehm_lattice_2012-1}.
Coupling between fluid and cell is achieved via the immersed-boundary algorithm \citep{Devendran_2012, saadat_immersed-finite-element_2018} which we implemented into ESPResSo \citep{Bacher_2017, Bacher_2019}.
\new{We note here that, in contrast to \citet{saadat_immersed-finite-element_2018}, we do not subtract the fluid stress within the particle interior.
This leads to a small viscous response of the cell material in addition to its elasticity. 
To obtain (approximately) the limit of a purely elastic particle, we exploit a recently developed method by \citet{Lehmann_2020} to discriminate between the cell interior and exterior during the simulation.
Using this technique, we can tune the ratio between inner and outer viscosity $ \lambda $ with $ \lambda \to 0$ representing a purely elastic particle.
For simplicity, we will nevertheless set $ \lambda =1$ in the following, except where otherwise noted.
}
Details of the method are provided in the SI (\sisref{4.1}).
As measure for the deformation, we investigate the Taylor parameter $\taylordeformation$~\eqref{eq:taylor-deformation-parameter-definition} of our initially spherical cell model in shear flow at different shear rates $\rateofshear$.
\subsection{Single cell simulation}
\label{sec:single-particle-sim}
\begin{figure}
	\hspace{2em}
	\includegraphics[width=0.75\linewidth]{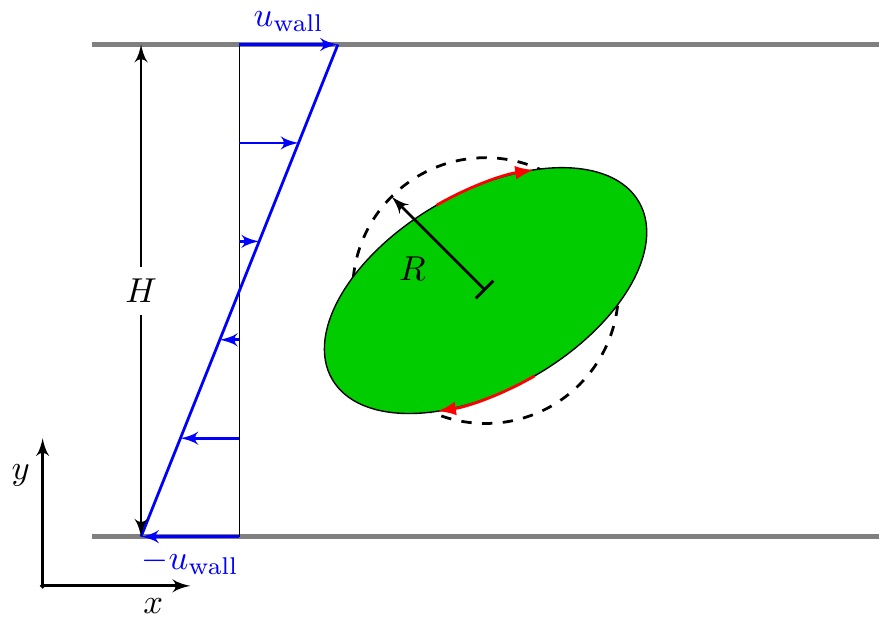}
	\caption{Schematic of the single cell in shear flow. The cell sits in the center of the box and shows an approximately elliptic deformation as well as tank-treading, \ie a rotation of the membrane around the steady shape in the $x$-$y$-plane
	}	
	\label{fig:singlecell-shear-simulation}
\end{figure}

\begin{figure}
	(a)\\
	\includegraphics[width=\linewidth]{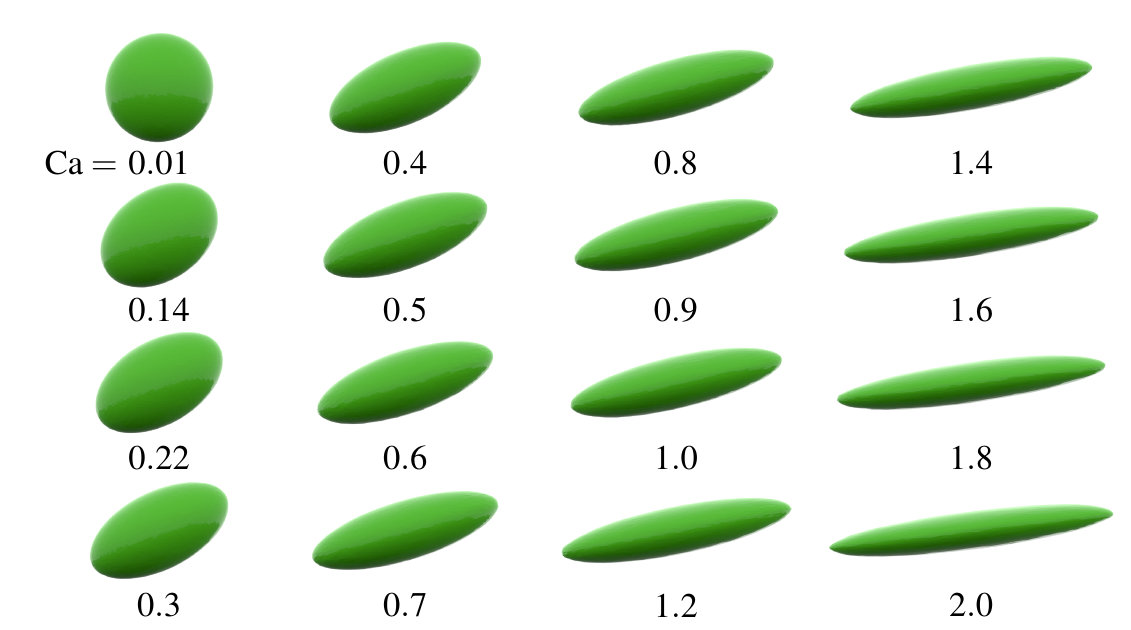}	
	(b)\\
	\includegraphics[width=\linewidth]{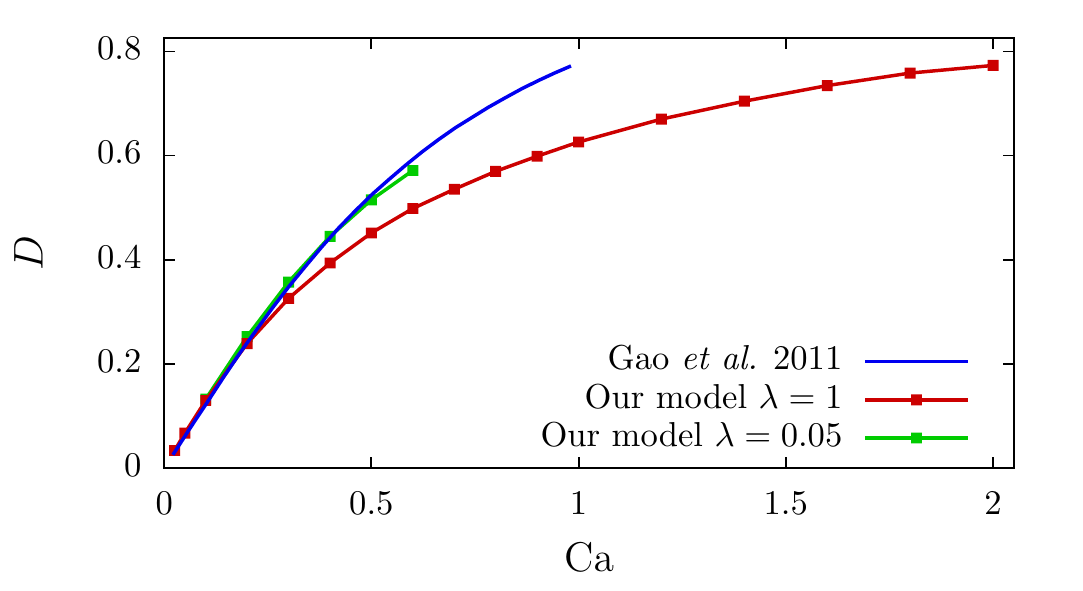}
	(c)\\
	\includegraphics[width=\linewidth]{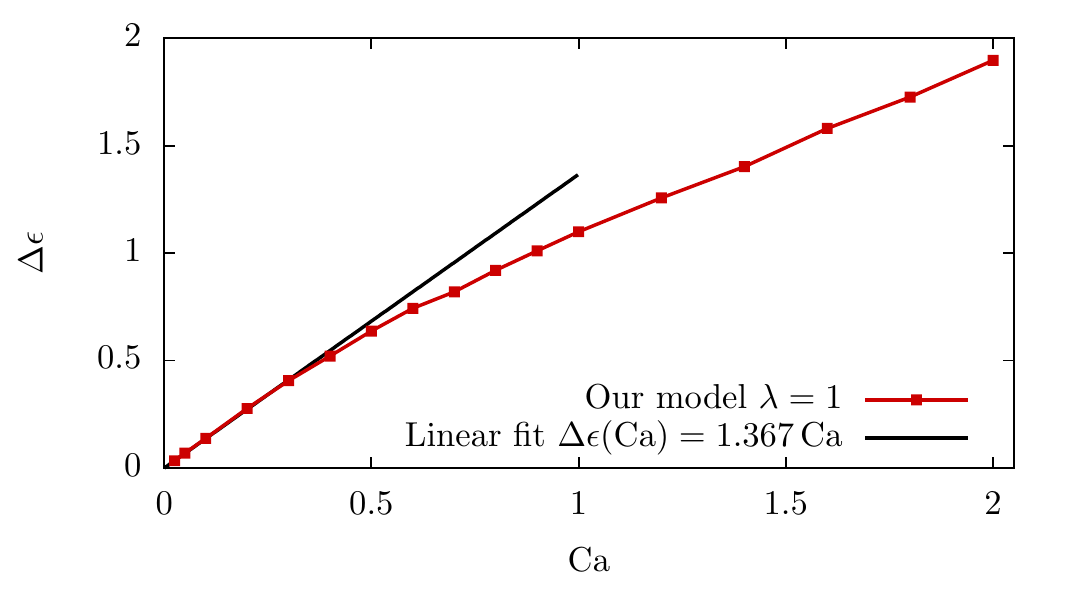}	
	\caption{
	(a) Converged shapes of a single cell in a $10\times15\times5$ ($x \times y \times z$) simulation box (in units of the cell radius) with a shear flow in $x$-direction as function of the capillary number $\capillarynumber$.
	(b) Comparison of our model predictions for a single cell in shear flow to the analytical 3D calculations in figure 7 of \citet{Gao_2011_JFM} in the range of $\capillarynumber \in \sbrak{0.01,2.0}$.
	(c) The relative stretch $\strain$ of our cell model as function of the capillary number $\capillarynumber$.
	A linear behavior is found for small capillary numbers up to $\capillarynumber=0.3$, while increasing stress is required for larger deformations due to the strain-hardening quality of the neo-Hookean model. Lines are a guide to the eye
	}
	\label{fig:taylor-deformation-comparison}	
\end{figure}

The first simulation setup, a single cell in infinite shear flow, is realized by choosing a simulation box of the dimensions $10\times15\times5$ ($x \times y \times z$) in units of the cell radius.
The infinite shear flow is approximated by applying a tangential velocity $\velocity\mindex{wall}$ on the $x$-$z$-planes at $y=0$ in negative and at $y=15$ in positive $x$-direction, as depicted in \fref{fig:singlecell-shear-simulation}.
The tangential wall velocity is calculated using the distance $H$ of the parallel planes and the constant shear rate $\rateofshear$ via
\begin{align}
\velocity\mindex{wall} = \half H \rateofshear \eqspace .
\end{align}
The box is periodic in $x$ and $z$.
A single cell is placed at the center of the simulation box corresponding to a volume fraction of $\volumefraction=0.0003$.
We choose the following parameters: fluid mass density $\density=\SI{e3}{\kilogram\per\cubic\meter}$, dynamic viscosity $\dynvisc=\SI{e-3}{\pascal\second}$, and shear rate $\rateofshear=\SI{4}{\per\second}$.
The capillary number is defined by \citep{gao_deformation_2009}
\begin{align}
\label{eq:capillary-number}
\capillarynumber = \frac{\dynvisc \rateofshear}{\shearmodulus} \eqspace ,
\end{align}
and is used to set the shear modulus $\shearmodulus$ of our cell relative to the fluid shear stress $\dynvisc \rateofshear$.
Simulation snapshots of the steady state deformation of a single cell in shear flow are depicted in dependency of the capillary number in \fref{fig:taylor-deformation-comparison}a.
We compare the Taylor deformation parameter $\taylordeformation$ to previous approximate analytical calculations of \citet{Gao_2011_JFM} for a three-dimensional elastic solid in infinite shear flow in \fref{fig:taylor-deformation-comparison}b and see reasonable agreement for our standard case of $ \lambda =1$.
\new{Reducing the inner viscosity by setting $ \lambda =0.05$, i.e. close to the limit of a purely elastic solid, the agreement is nearly perfect.
Finally, we demonstrate that the elastic particle exhibits a tank-treading motion in \sisref{4.2}.}

A possibly even more intuitive way to measure cell deformation is the net strain of the cell which we define as
\begin{align}
\label{eq:cell-strain}
\strain = \frac{\rbrak{d\mindex{max}-d\mindex{ref}}}{d\mindex{ref}} \eqspace .
\end{align}
It describes the relative stretching of the cell using the maximum elongation $d\mindex{max}$, \ie the maximum distance of two cell vertices, and its reference diameter $d\mindex{ref}=2\cellradius$.
A strain of $\strain=\num{1}$ thus corresponds to an elongation of the cell by an additional $\SI{100}{\percent}$ of its original size.
In \fref{fig:taylor-deformation-comparison}c, we depict the $\strain$ as function of $\capillarynumber$.
For small capillary numbers, \ie small shear stresses, a linear stress-strain dependency is observed.
Above $\capillarynumber\approx 0.3$, the strain-hardening, nonlinear behavior of the neo-Hookean model can be seen.
By stretching the cell up to $\SI{280}{\percent}$ of its initial size, this plot demonstrates again the capability of our model to smoothly treat large deformations.
\subsection{Multiple cell simulations}
\label{sec:multiple-particle-sim}
The second simulation setup, implemented to investigate the multiple particle aspect of our model, consists of $4$ ($8$) cells in a $5\times8\times4$ simulation box (in units of the cell radius), corresponding to a volume fraction of $\volumefraction=0.11$ ($\volumefraction=0.22$) occupied by cells.
The cells are inserted at random initial positions in the box and the flow parameters are the same as in the first setup (cf.~\sref{sec:single-particle-sim}).

\begin{figure}
	(a)\\
	\includegraphics[width=\linewidth]{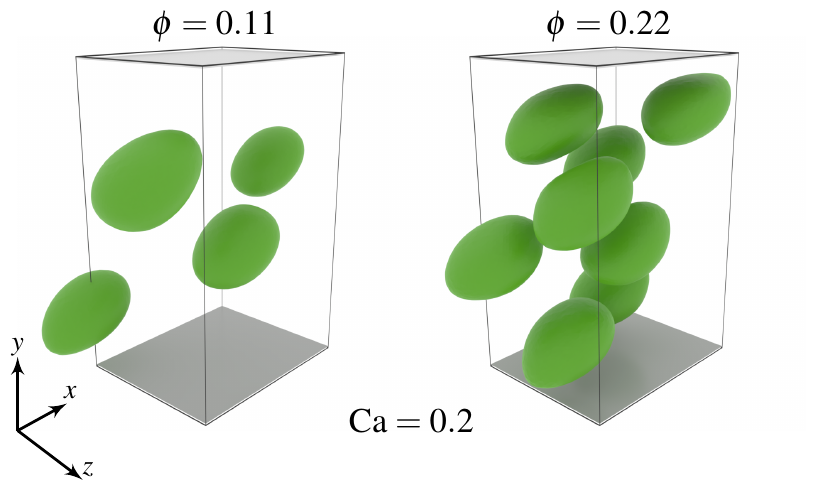}	
	(b)\\
	\includegraphics[width=\linewidth]{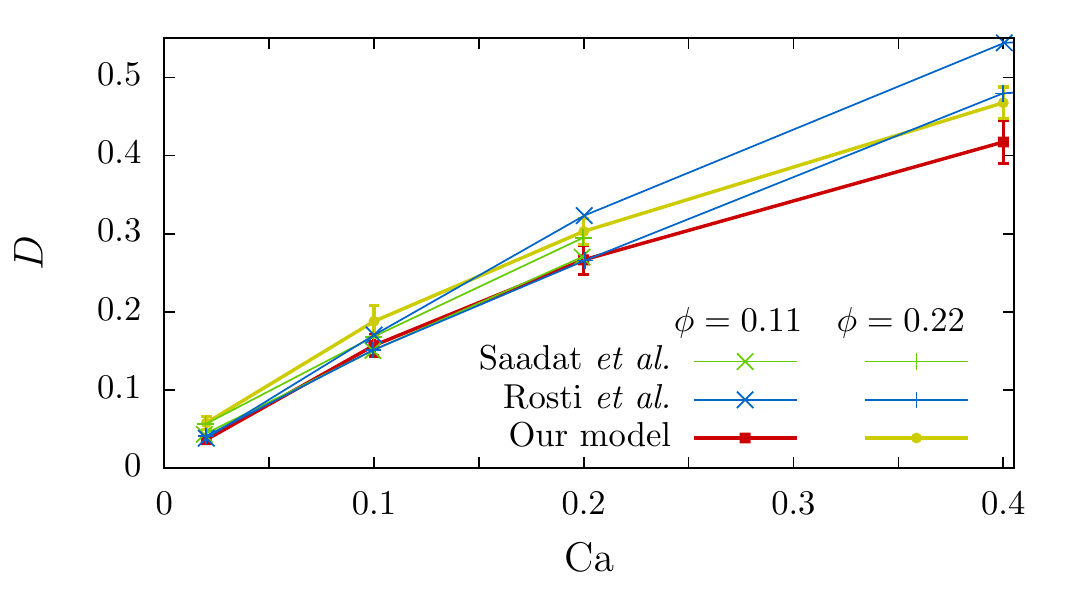}
	\caption{
		(a) Multiple cells in a $5\times8\times4$ ($x \times y \times z$) simulation box (in units of the cell radius) with a confined shear flow in $x$-direction for a capillary number of $\capillarynumber=0.2$ and $4$ cells corresponding to a volume fraction of $\volumefraction=0.11$, and $8$ cells corresponding to $\volumefraction=0.22$.
\new{		(b) Averaged deformation of multiple cell simulations with $\volumefraction=0.11$ and $\volumefraction=0.22$ in comparison to data from figure 3 of \citet{rosti_rheology_2018} and figure 13 of \citet{saadat_immersed-finite-element_2018}}
	}
	\label{fig:multicell-shear-simulation}		
\end{figure}

\Fref{fig:multicell-shear-simulation}a shows simulation snapshots of the cells in suspensions with volume fraction $\volumefraction=0.11$ and $\volumefraction=0.22$ for $\capillarynumber=0.2$.
The Taylor deformation of the suspensions, depicted in \fref{fig:multicell-shear-simulation}b, is calculated as an average over all cells and over time after an initial transient timespan.
We find good agreement when comparing the averaged cell deformation in suspension with \citet{rosti_rheology_2018,saadat_immersed-finite-element_2018}.
%
\section{Conclusion}
We presented a simple but accurate numerical model for cells and other microscopic particles for the use in computational fluid-particle dynamics simulations.

The elastic behavior of the cells is modeled by applying Mooney--Rivlin strain energy calculations on a uniformly tetrahedralized spherical mesh.
We performed a series of FluidFM\textsuperscript{\textregistered}\  compression experiments with REF52 cells as an example for cells used in bioprinting processes and found excellent agreement between our numerical model and the measurements \new{if all three parameters of the Mooney--Rivlin model are used}.
In addition, we showed that the model compares very favorably to force versus deformation data from previous AFM compression experiments on bovine endothelial cells \citep{caille_contribution_2002-1} as well as colloidal probe AFM indentation of artificial hydrogel particles \citep{neubauer_mechanoresponsive_2019-1}.
At large deformations, a clear improvement compared to Hertzian contact theory has been observed.	

By coupling our model to Lattice Boltzmann fluid calculations via the Immersed-Boundary method, the cell deformation in linear shear flow as function of the capillary number was found in good agreement with analytical calculations by \citet{Gao_2011_JFM} on isolated cells as well as previous simulations of neo-Hookean and viscoelastic solids \citep{rosti_rheology_2018,saadat_immersed-finite-element_2018} at various volume fractions.

The presented method together with the precise determination of model parameters by FluidFM\textsuperscript{\textregistered}\ /AFM experiments may provide an improved set of tools to predict cell deformation - and ultimately cell viability - in strong hydrodynamic flows as occurring, e.g., in bioprinting applications.
%
%
\section*{Acknowledgements}
Funded by the Deutsche Forschungsgemeinschaft (DFG, German Research Foundation) --- Project number 326998133 --- TRR 225 ``Biofabrication'' (subproject B07). 
We gratefully acknowledge computing time provided by the SuperMUC system of the Leibniz Rechenzentrum, Garching.
We further acknowledge support through the computational resources provided by the Bavarian Polymer Institute.
Christian B\"acher thanks the Studienstiftung des deutschen Volkes for financial support and acknowledges support by the study program ``Biological Physics'' of the Elite Network of Bavaria.
Furthermore, we thank the laboratory of professor Alexander Bershadsky at Weizmann Insitute of Science in Isreal for providing the REF52 cells stably expressing paxillin-YFP.
%
%
%

%
\cleardoublepage


\section*{Supplementary material}

## S-1 Supplementary Material for the cell experiments

Additional force-deformation curves for our FluidFM® measurements on REF52 cells are shown in figure S-1. Compared to the curves depicted in the manuscript in figure 4, these measurements show an earlier upturn of the force. Thus, our model overestimates the force necessary for a small deformation of the cell and slightly underestimates the force for larger deformations. Nevertheless, all measurements fit in the simulated range of $E = 220 \pm 100 \, \mathrm{Pa}$ for $w = 0.25$ and an averaged cell radius of $8.6(7) \, \mathrm{\mu m}$, as figure S-1 shows. The cell radii and Young's moduli for all measurements are listed in table S-1.

**Table S-1** Measured cell radii $R$ and fitted Young's moduli $E$ and $w$ for our FluidFM® experiments.

| Number | 1 | 2 | 3 | 4 | 5 | 6 | 7 | 8 | 9 |
|---|---|---|---|---|---|---|---|---|---|
| $R$ [μm] | 7.1 | 9.2 | 8.3 | 8.0 | 9.5 | 9.1 | 8.4 | 9.4 | 8.3 |
| $E$ [Pa] | 160 | 190 | 220 | 170 | 210 | 290 | 210 | 220 | 125 |
| $w$ | 1 | 0.25 | 0.25 | 0.5 | 0.25 | 0.25 | 0.25 | 0.25 | 0.25 |

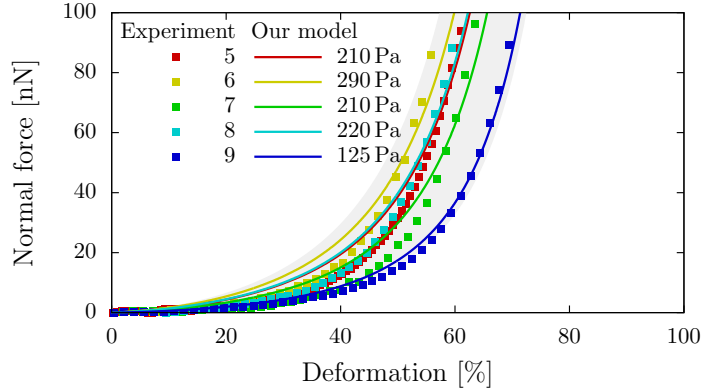

**Fig. S-1** Our numerical model in comparison to our FluidFM® measurements on REF52 cells. The ratio of the shear moduli is chosen as $w = 0.25$ for all curves. The gray area shows the simulation of a cell with an averaged cell radius of $8.6(7) \, \mathrm{\mu m}$ and Young's modulus range $220 \pm 100 \, \mathrm{Pa}$.

## S-2 Supporting Information for the numerical model

### S-2.1 Convergence of single cell deformation in shear flow

The temporal development of the deformation $D$ of a single cell in a Couette flow can be seen in figure S-2. Starting from a spherical shape ($D = 0$), the cell experiences a shape change during an initial transient timespan, after which it assumes a steady shape. For capillary numbers Ca > 0.2, we first find an overrelaxation of the deformation before it converges towards a constant value.

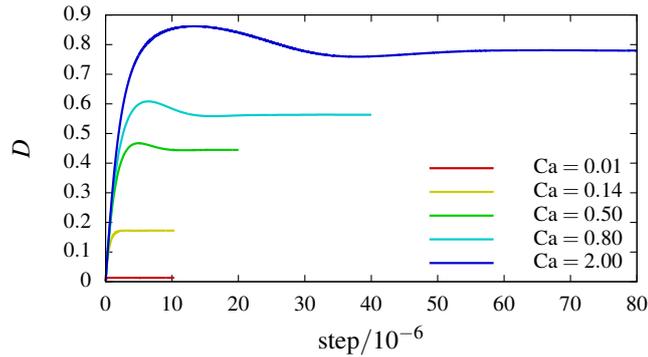

**Fig. S-2** Single cell deformation in Couette flow for different capillary numbers. After an initial transient timespan, the deformation converges to a constant value.

### S-2.2 Reduction of the system resolution

In figure S-3 we show that a system with reduced cell resolution (from $R_{\mathrm{Cell}} = 10$ to $R_{\mathrm{Cell}} = 6$ grid cells) and a smaller simulation box (from $100 \times 150 \times 100$ to $60 \times 90 \times 30$ grid cells) produces the same deformation versus capillary number behavior as the system with higher resolution.

### S-2.3 Translational and rotational invariance of the force calculation

As a very direct test for the correct behavior of our model, we consider a single tetrahedron and examine the behavior of the volume and the elastic force for an initially applied translation, rotation and stretching. In figure S-4a, the behavior of the volume under these deformations is shown over the first time steps. While the volume remains constant under pure translation, pure rotation, and a combination of both, it quickly relaxes towards its reference value after an initial stretch is applied. The same behavior is observed for the elastic force acting on one tetrahedron vertex, in figure S-4b.

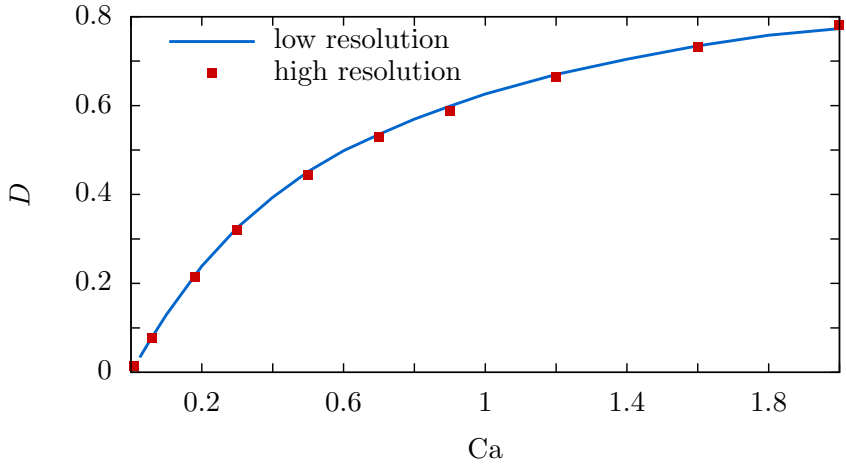

**Fig. S-3** Taylor deformation as function of the capillary number for two different cell and channel resolutions. The large system ($R_{\text{Cell}} = 10$, box: $100 \times 150 \times 100$ grid cells) produces the same outcome as the down-scaled system ($R_{\text{Cell}} = 6$, box: $60 \times 90 \times 30$ grid cells).

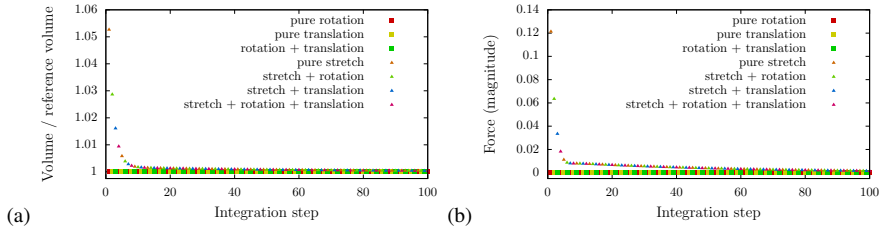

**Fig. S-4** The behavior of (a) the volume and (b) the elastic force on a single vertex of a tetrahedron after an initial rotation, translation or stretching.

## S-2.4 Mesh generation and mesh independence

The tetrahedral mesh of our spheroid is generated using the software gmsh (version 4.3.0) [1]. The Frontal2D meshing algorithm produced a mesh with highest uniformity considering edge length, triangle area and tetrahedron volume distribution. Nevertheless, all other available meshing algorithms produce likewise uniform meshes, with one exception being the Frontal3D algorithm, as listed in table S-2. We demand the uniformity of the mesh to increase the accuracy of our coupled Immersed-Boundary Lattice Boltzmann simulations. Figure S-5 shows the force-deformation curves for meshes with increasing number of tetrahedra, which are converged and thus prove sufficient sampling of the volume mesh.

**Table S-2** Statistics of meshes created using different built-in algorithms of Gmsh [1]. Listed are edge length $L$, triangle area $A$, and tetrahedron volume $V$ providing average, standard deviation, minimum and maximum value for each mesh.

| Algorithm | Frontal2D | MeshAdapt | Delaunay2D | Delaunay3D | Frontal3D |
|---|---|---|---|---|---|
| $\bar{L}$ | 1.252 | 1.362 | 1.292 | 1.362 | 1.484 |
| $\sigma_L$ | 0.243 | 0.301 | 0.299 | 0.301 | 0.530 |
| $L_{min}$ | 0.616 | 0.588 | 0.592 | 0.588 | 0.510 |
| $L_{max}$ | 2.138 | 2.345 | 2.462 | 2.345 | 3.622 |
| $\bar{A}$ | 0.348 | 0.422 | 0.382 | 0.422 | 0.565 |
| $\sigma_A$ | 0.377 | 0.473 | 0.436 | 0.473 | 0.837 |
| $A_{min}$ | 0.218 | 0.228 | 0.192 | 0.228 | 0.204 |
| $A_{max}$ | 1.577 | 1.851 | 1.709 | 1.851 | 4.444 |
| $\bar{V}$ | 0.218 | 0.291 | 0.252 | 0.291 | 0.473 |
| $\sigma_V$ | 0.078 | 0.121 | 0.112 | 0.121 | 0.405 |
| $V_{min}$ | 0.049 | 0.051 | 0.043 | 0.051 | 0.049 |
| $V_{max}$ | 0.600 | 0.881 | 0.840 | 0.881 | 2.353 |

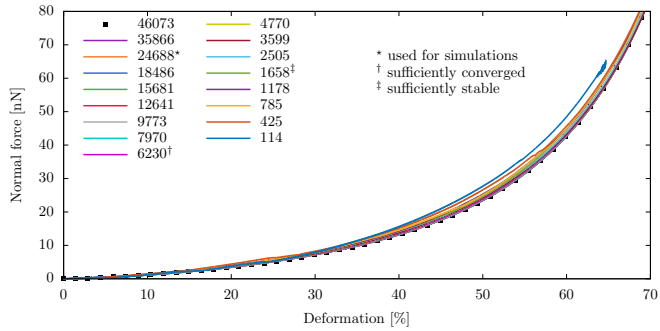

**Fig. S-5** Force-deformation behavior of meshes with increasing number of tetrahedra. Meshes with $N \geq 1658$ tetrahedra are stable in the investigated range of deformation. Above 6230 tetrahedra, all meshes produce the same converged output. The following parameters were used: cell radius $R = 7.5\,\mu m$, Young's modulus $E = 300\,Pa$, and Poisson ratio $\nu = 0.48$.

### S-2.5 Hertz theory

Although originally designed for the contact between two linear elastic spheres, the Hertz theory can be applied to the contact between a linear elastic sphere and a flat plate [2]. The general assumptions for the Hertz-theory are the following [3, p. 91-92]:

- frictionless, smooth contact surfaces
- contact area small compared to sphere dimension
- homogeneous, isotropic and linear elastic material

#### S-2.5.1 Sphere-sphere contact

The following quantities are necessary to describe the normal contact of two elastic spheres. The radii $R_1$ and $R_2$ of the spheres define the effective radius of curvature $R$

of the bodies by

$$\frac{1}{R} = \frac{1}{R_1} + \frac{1}{R_2} \quad . \tag{S-1}$$

Through their Young's moduli and the Poisson ratios, $E_1, E_2$ and $\nu_1, \nu_2$, the effective stiffness $K$ is defined as:

$$\frac{1}{K} = \frac{1 - \nu_1^2}{E_1} + \frac{1 - \nu_2^2}{E_2} \tag{S-2}$$

The displacement $\delta$, which measures the distance that the sphere centers approach each other due to a normal force $N$ acting on each sphere, can be expressed in terms of the above parameters [2]:

$$\delta = \left( \frac{9N^2}{16RK^2} \right)^{\frac{1}{3}} \tag{S-3}$$

Therefore, the force–displacement relation according to the Hertzian theory for a sphere-sphere contact is given by

$$N(\delta) = \frac{4}{3} K R^{\frac{1}{2}} \delta^{\frac{3}{2}} \quad . \tag{S-4}$$

### S-2.5.2 Sphere-plane contact

The analytical solution for the force–displacement relation according to the Hertzian theory for the contact of a linear elastic sphere with a rigid plane can be obtained from (S-4) by applying the following modifications: the plane has no curvature, thus $R_2 \to \infty$ and (S-1) simply yields $R = R_1$. Since the plane is assumed rigid, i. e. $E_2 \gg E_1$, (S-2) reduces to $K = \frac{E_1}{1 - \nu_1^2}$. In this case, $N$ is the force acting on the sphere and $\delta$ is the distance between the center of the sphere and the plane.

### S-2.6 Influence of the Mooney-Rivlin ratio $w$

To clarify the influence of $w$, we plot in figure S-6 the force versus deformation behavior of our cell model for different values of $w$. With decreasing $w$, i. e. decreasing $\mu_1$ while increasing $\mu_2$, the strain hardening effect clearly increases and the upturn of the force curve begins at lower deformations. This is due to $\mu_2$ scaling the term in the strain energy density that is quadratic with the deformation (cf. equations (4) and (5) of the manuscript).

### S-2.7 Influence of the Poisson ratio $\nu$

In figure S-7 we demonstrate that variations of the Poisson ratio $\nu$ within the range of an approximately incompressible material do not notably influence the force-deformation curves.

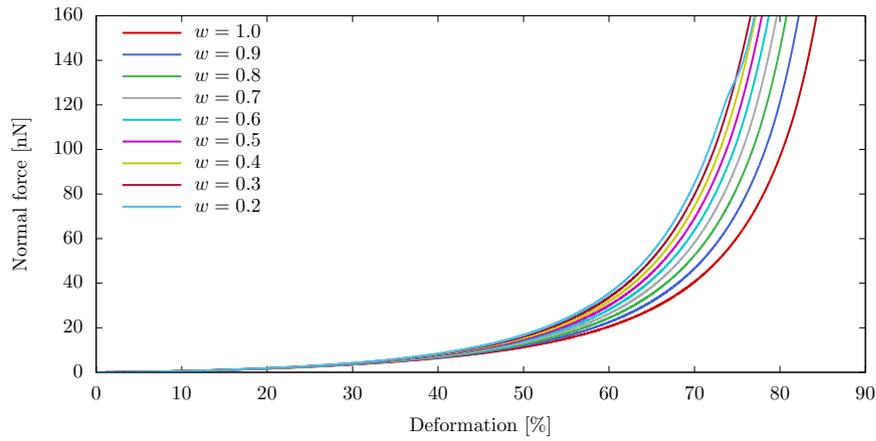

**Fig. S-6** Variation of $w$: The lower $w$ (the lower $\mu_1$ compared to $\mu_2$), the stronger the non-linear upturn of the force becomes. The curve with $w = 1$ corresponds to the one in Fig. 4 of the manuscript with a Young's modulus of $E = 160$ Pa.

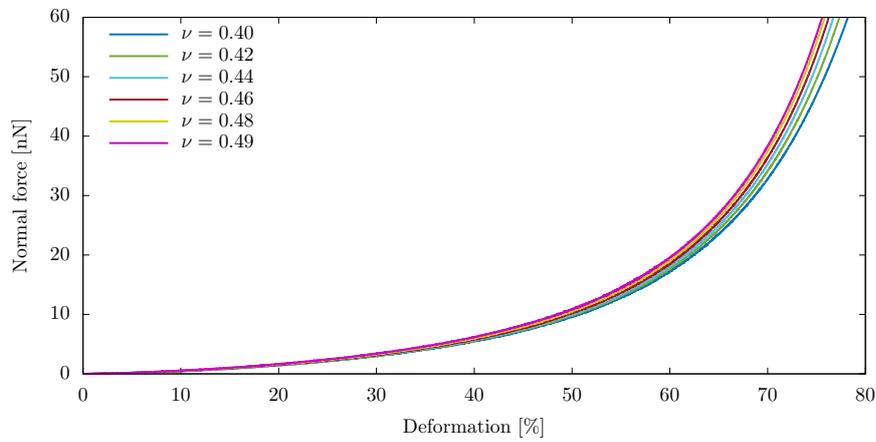

**Fig. S-7** Force versus deformation curves for different Poisson ratios $\nu$. The following parameters were used: cell radius $R = 7.5\,\mu$m and Young's modulus $E = 300$ Pa.

## S-3 Compression and indentation simulations

After initialization, each time step of our overdamped relaxation simulation consists of the following two steps: the movement of the upper wall to compress – or the sphere to indent – the cell and the integration of the equation of motion of the cell vertices,

$$\dot{\mathbf{y}}^{\alpha} = \gamma^{-1}(\mathbf{f}^{\alpha} + \mathbf{f}^{\alpha}_{\text{probe}}) \quad . \tag{S-5}$$

The vertex velocity $\dot{\mathbf{y}}^{\alpha}$ is obtained from the elastic restoring forces ($\mathbf{f}^{\alpha}$ (12) and the probe repulsion $\mathbf{f}^{\alpha}_{\text{probe}}$), considering a friction factor $\gamma$. Since here we are only looking at a sequence of equilibrium states, the value of $\gamma$ is irrelevant for the resulting force-deformation curves and only influences the performance and stability of the simulations. The equation of motion is integrated using a fourth order Runge-Kutta algorithm. The repulsive cell-probe interaction, preventing the cell vertices from penetrating the plates or the indenter, has the form

$$\mathbf{f}_{\text{probe}}(d) = \frac{c_{\text{F}}}{d^2}\mathbf{n} \quad , \tag{S-6}$$

with the cell-probe distance $d$ and a proportionality factor $c_{\text{F}}$. The force points normal to the probe, resulting in a compression between two plates and a radial displacement away from the indenter. Physically, this corresponds to a free-slip boundary condition which does not restrict tangential motions of the cell along the probe.

**S-4 Flow simulations with Lattice Boltzmann**

S-4.1 Method

This section briefly summarizes the Lattice Boltzmann method implemented in the open-source package ESPResSo [4]. For an introduction into the Lattice Boltzmann method we refer the interested reader to the book by Krüger et al. [5]. The Lattice Boltzmann equation for the multiple relaxation time scheme used in ESPResSo reads:

$$f_i(\mathbf{x} + \mathbf{c}_i \Delta t, t + \Delta t) - f_i(\mathbf{x}, t) = \sum_{j=0}^{18} \left(\mathsf{M}^{-1} \boldsymbol{\omega} \mathsf{M}\right)_{ij} \left(f_j(\mathbf{x}, t) - f_j^{\text{eq}}(\mathbf{x}, t)\right) \qquad \text{(S-7)}$$

It describes the collision and streaming of the population distribution $f_i$ ($i = 0, \ldots, 18$) during one time step $\Delta t$. Here, $\mathbf{c}_i$ are the discretized lattice velocities, $\mathsf{M}$ denotes transformation matrix that maps the populations onto moment space, $\boldsymbol{\omega}$ is the diagonal relaxation frequency matrix, and $f_i^{\text{eq}}$ denote the equilibrium population distributions. The relaxation frequency for the shear moments $\omega_{\text{S}}$ is related to the dynamic viscosity of the fluid via [6]

$$\eta = \rho c_s^2 \left(\frac{1}{\omega_{\text{S}}} - \frac{1}{2}\right) \Delta t \quad , \qquad\qquad \text{(S-8)}$$

with the fluid mass density $\rho$ and the lattice speed of sound $c_s$. In order to ensure simulation stability, we choose the time step globally according to Krüger *et al.*[5, p. 273] as

$$\Delta t = c_s^2 \left(\tau - \frac{1}{2}\right) \frac{\Delta x^2}{\nu} \tilde{t} = \frac{\Delta x^2}{6\nu} \tilde{t} \quad , \qquad\qquad \text{(S-9)}$$

with $c_s^2 = \frac{1}{3}$, a global relaxation parameter $\tau = 1$, the kinematic viscosity $\nu$, and an additional factor $\tilde{t}$ in the range 1–2 to manually tune the time step.

We further introduce a scaling factor $r$ by which we divide both the viscosity and the Young's modulus. According to eq. (S-9), this leads to a larger time step and thus to a speed-up of the simulations. At the same time it leaves the important Capillary number unchanged and only increases the Reynolds number, which nevertheless remains $\ll 1$. The parameter $r$ thus does not affect the physics of the simulation which we have carefully checked by a number of test runs with $r = 1$.

At the boundaries of the channel a bounce-back algorithm is applied to realize a no-slip boundary condition. For the plane Couette setup, the bounce-back algorithm additionally allows for a fixed tangential velocity component.

We use a combined CPU/GPU implementation which enables the calculation of the flow field on the GPU, while the calculation of the cell motion is done in parallel on multiple (4 to 20) CPUs. In lattice units, our simulation box for the single cell in shear flow setup (cf. section 6.1) has the dimensions $60 \times 90 \times 30$ ($x \times y \times z$), for the multiple cell simulation (cf. section 6.2) it is $50 \times 80 \times 40$. The dynamic viscosity, chosen as $\nu = 1$ in simulation units, determines the time step in our simulations as $\Delta t = \frac{1}{3}$ with $\tilde{t} = 2$.

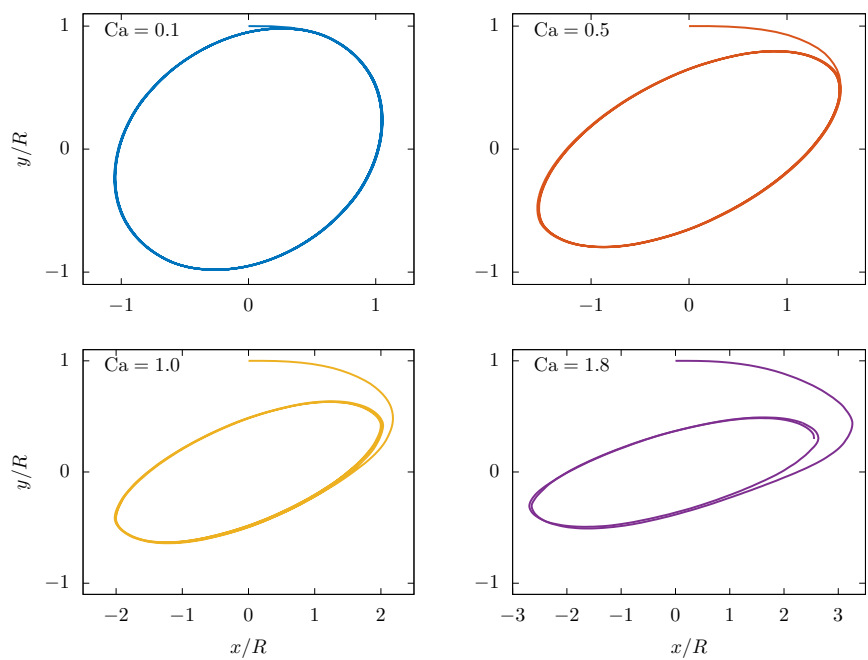

**Fig. S-8** The trajectory of a surface node (here: starting at $y = R$ and $x, z = 0$) for different capillary numbers traces the ellipsoidal contour of the deformed particle. The non-elliptical part of the trajectory in the upper-right corner represents the approach from the initially spherical to the final shape.

### S-4.2 Tank-treading motion

Figure S-8 shows the trajectories of selected vertices on the outer surface of the particle for different capillary numbers. They describe an ellipsoidal motion tracing the outer contour of the deformed particle thus demonstrating that in our simulations the particle exhibits tank-treading.


\end{document}